\documentclass{egpubl}
\usepackage{egsr18}

\usepackage{amsmath,amsfonts,amssymb,bm}
\usepackage[usenames,dvipsnames]{color}
\usepackage{graphicx,adjustbox}
\usepackage[ruled,vlined]{algorithm2e} 
\usepackage{url}
\usepackage{booktabs}
\usepackage{subcaption}
\usepackage{float}
\usepackage{multirow}
\usepackage{changepage}
\usepackage{float}

\usepackage{overpic,transparent,multirow,varwidth,wrapfig}
\usepackage{url}

\makeatletter
\newcommand{\removelatexerror}{\let\@latex@error\@gobble}
\makeatother

\definecolor{darkgreen}{RGB}{10,150,10}
\definecolor{gray}{RGB}{128,128,128}
\newcommand{\es}[1] {\emph{\textcolor{darkgreen}{ES: #1}}}
\newcommand{\kb}[1] {\emph{\textcolor{blue}{KB: #1}}}

\newcommand{\fujun}[1] {{{#1}}}
\newcommand{\ignorethis } [1] {}
\renewcommand{\paragraph} [1] {\noindent\textbf{#1}}

\newcommand{\img}{I}
\newcommand{\mask}{M}
\newcommand{\style}{S}
\newcommand{\out}{O}
\newcommand{\inter}{I'}
\newcommand{\neuralcoef}{F}
\newcommand{\neuralmapping}{\ensuremath{\pi}}
\newcommand{\stylemap}{P}
\newcommand{\stylemapout}{\stylemap_\text{\rm out}}

\newcommand{\neurallayer}{\ell}
\newcommand{\refneurallayer}{{\neurallayer_\text{\rm ref}}}
\newcommand{\gram}{G}

\newcommand{\neuralfilter}{i}
\newcommand{\neuralfilterp}{j}
\newcommand{\neuralpixel}{p}
\newcommand{\neuralpatch}{p}
\newcommand{\candidatepatch}{c}
\newcommand{\neuralstylepatch}{q}
\newcommand{\refneuralpatch}{\neuralpatch'}

\newcommand{\loss}{\mathcal{L}}
\newcommand{\Gatysloss}{\loss_\text{\rm Gatys}}
\newcommand{\contentloss}{\loss_\text{\rm c}}
\newcommand{\styleloss}{\loss_\text{\rm s}}
\newcommand{\uniquestyleloss}{\loss_{\text{\rm s1}}}
\newcommand{\histloss}{\loss_\text{\rm hist}}

\newcommand{\tvloss}{\loss_\text{\rm tv}}
\newcommand{\finalloss}{\loss_\text{\rm final}}

\newcommand{\weight}{w}
\newcommand{\styleweight}{\weight_\text{\rm s}}
\newcommand{\histweight}{\weight_\text{\rm hist}}
\newcommand{\tvweight}{\weight_\text{\rm tv}}

\newcommand{\numberoflayers}{L}
\newcommand{\numberoffilters}{N}
\newcommand{\numberofcoefs}{D}
\newcommand{\numberofpatches}{D}

\newcommand{\linecomment}[1]{{\small\textcolor{gray}{\tcp{#1}}}}
\newcommand{\eolcomment}[1]{{{\small\color{gray}\tcp*[h]{#1}}}}

\newcommand{\transpose}[1]{#1^\mathsf{T}}

\newcommand{\histmatch}{R}

\newcommand{\offset}{o}
\newcommand{\adjoffsetsetshort}{\mathcal{O}}
\newcommand{\adjoffsetset}{\{\mathit{N},\mathit{NE},\mathit{E},\mathit{SE},\mathit{S},\mathit{SW},\mathit{W},\mathit{NW}\}}

\DeclareMathOperator{\argmin}{arg\,min}

\newcommand{\vgglayer}[1]{\textsf{\textsl{#1}}}

\SpecialIssuePaper         
 \electronicVersion 

\PrintedOrElectronic

\usepackage{t1enc,dfadobe}

\usepackage{egweblnk}
\usepackage{cite}

\let\rm=\rmfamily        
\let\it=\itshape          
\let\bf=\bfseries

\title{Deep Painterly Harmonization}

\author[Luan et al.]
{\parbox{\textwidth}{\centering Fujun Luan$^{1}$ \quad
        Sylvain Paris$^{2}$ \quad
        Eli Shechtman$^{2}$ \quad
        Kavita Bala$^{1}$ \quad
        }
        \\
{\parbox{\textwidth}{\centering $^1$Cornell University  \quad
         $^2$Adobe Research 
       }
}
}

\begin{document}

\teaser{
 \includegraphics[width=\linewidth]{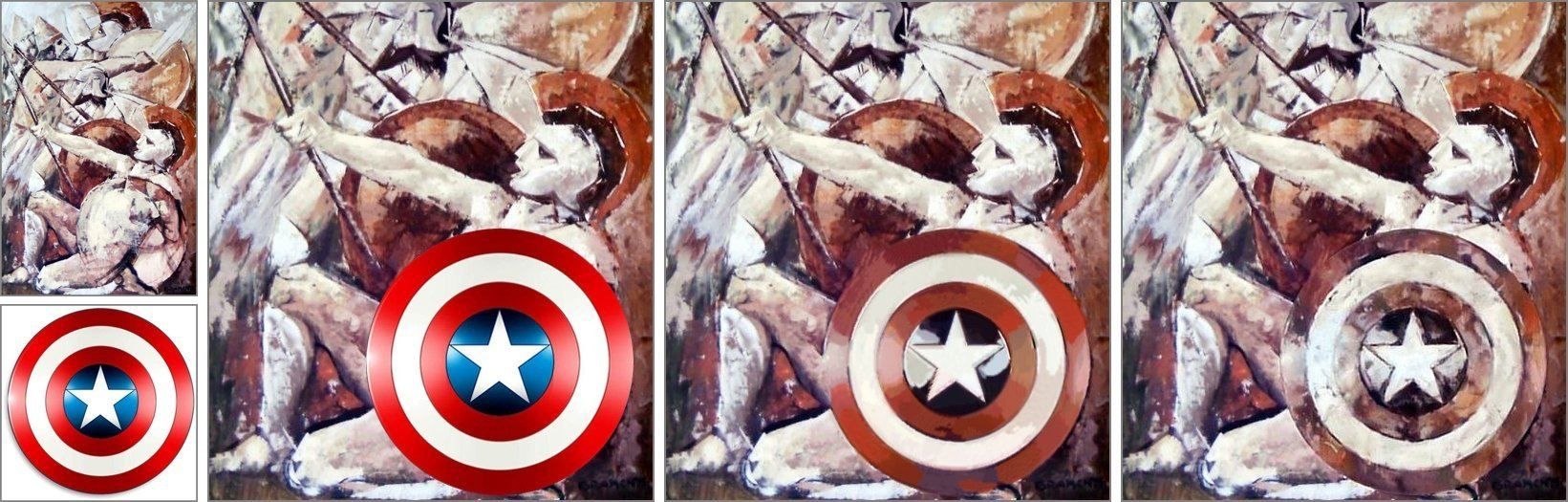}
 \centering
  \caption{Our method automatically harmonizes the compositing of an element into a painting. Given the proposed painting and element on the left, we show the compositing results (cropped for best fit) of unadjusted cut-and-paste, Deep \fujun{Image} Analogy~\protect\cite{liao2017visual}, and our method.}
\label{fig:teaser}
}

\maketitle
\begin{abstract}
    Copying an element from a photo and pasting it into a painting is a
    challenging task. Applying  photo compositing techniques in this
    context yields subpar results that look like a collage --- and
    existing painterly stylization algorithms, which are global, perform
    poorly when applied locally. We address these issues with a
    dedicated algorithm that carefully determines the local statistics
    to be transferred. We ensure both spatial and inter-scale
    statistical consistency and demonstrate that both aspects are key to
    generating quality results. To cope with the diversity of
    abstraction levels and types of paintings, we introduce a technique
    to adjust the parameters of the transfer depending on the
    painting. We show that our algorithm produces significantly better
    results than photo compositing or global stylization techniques and
    that it enables creative painterly edits that would be otherwise
    difficult to achieve.


\end{abstract}  
  \section{Introduction}

Image compositing is a key operation to create new visual content. It
allows artists to remix existing materials into new pieces and
artists such as Man Ray and David Hockney have created masterpieces
using this technique. Compositing can be used in different contexts.
In applications like photo collage, visible seams are desirable.
But in others, the objective is to make the compositing inconspicuous,
for instance, to add an object into a photograph in a way that makes
it look like the object was present in the original scene. Many tools
have been developed for photographic compositing, e.g., to remove
boundary seams~\cite{perez2003poisson}, match the color~\cite{xue2012understanding} or also fine texture~\cite{sunkavalli2010multi}. However, there is no
equivalent for paintings. If one seeks to add an object into a
painting, the options are limited. One can paint the object manually
or with a painting engine~\cite{chen2015wetbrush} but this requires time and
skills that few people have. As we shall see, resorting to algorithms designed
for photographs produces subpar results because they do not handle the brush
texture and abstraction typical of paintings. And applying existing painterly
stylization algorithms as is also performs poorly because they are meant for
global stylization whereas we seek a local harmonization of color, texture, and
structure properties. 

In this paper, we address these challenges and enable one to copy an
object in a photo and paste it into a painting so that the composite
still looks like a genuine painting in the style of the original
painting. We build upon recent work on painterly
stylization~\cite{gatys2015neural} to harmonize the appearance of the
pasted object so that it matches that of the painting.  Our strategy
is to transfer {\em relevant} statistics of neural responses from the
painting to the pasted object, with the main contribution being how we
determine which statistics to transfer. Akin to previous work, we use
the responses of the VGG neural network~\cite{simonyan2014very} for
the statistics that drive the process. In this context, we show that
spatial consistency and inter-scale consistency matter. That is,
transferring statistics that come from a small set of regions in the
painting yields better results than using many isolated locations.
Further, preserving the correlation of the neural responses between
the layers of the network also improves the output quality.  To
achieve these two objectives, we introduce a two-pass algorithm: the
first pass achieves coarse harmonization at a single scale. This
serves as a starting point for the second pass which implements a fine
multi-scale refinement. Figure~\ref{fig:teaser}(right) shows the
results from our approach compared to a related technique.

We demonstrate our approach on a variety of examples.  Painterly compositing is
a demanding task because the synthesized style is juxtaposed with the
original painting, making any discrepancy immediately visible. As a
consequence, results from global stylization techniques that may be
satisfying when observed in isolation can be disappointing in the context
of compositing because the inherent side-by-side comparison with the
original painting makes it easy  to identify even subtle
differences. In contrast, we conducted a user study that shows that
our algorithm produces composites that are often perceived as genuine
paintings.

\ignorethis{
	\hrule
	
	Paintings often exhibit unique visual appearances through the composing of brush strokes. By changing the type of brushes and the techniques to draw strokes, artists can create various styles by interplaying the color and texture, yielding different perception experiences of the audience. The task of copying an element from a photo and pasting it into a painting falls into the category of image compositing. To generate such a composite image, the foreground is extracted from one image and combined with the background of another image. However, the foreground and background may have different appearances such as inconsistent color or texture, making the composite look unnatural. Therefore, it is crucial to adjust the appearance of the foreground element and harmonize it with the background painting, as shown in Figure~\ref{fig:teaser}.
	
	In this work, we introduce a novel algorithm to transfer both spatial and inter-scale color and texture from the background painting to the foreground element semantically. The main contributions include:
	\begin{itemize}
		
		\item A \emph{painting harmonizer} that selects neural patches of a painting by computing the correspondence field in deep neural space and synthesizes the compositing output through optimization;
		
		\item An optional \emph{painting estimator} that estimates the level of stylization on a painting and predicts optimization parameters for the painting harmonizer accordingly.
		
	\end{itemize}
}

\ignorethis{
	Image compositing is a fundamental and challenging problem in image editing. Given the proposed foreground and background, unadjusted cut-and-paste often looks unnatural because of the color and texture inconsistency. Poisson editing~\cite{perez2003poisson} produces plausible boundaries by considering the gradient-domain interpolation. Multi-scale image harmonization~\cite{sunkavalli2010multi} further improves the quality by smooth histogram matching on decomposed pyramid layers and a noise simulator which mimics the background noise to better preserve the texture. However, it fails when the background painting is strongly stylized with noticeable brush strokes, as shown in Figure~\ref{fig:teaser}. 
	
	In this paper, we propose a deep-learning approach builds upon the recent work of neural style transfer~\cite{gatys2015neural} that separates content and style of an image via the statistics of neural network layers. Neural style transfer has shown its ability to mimic an artwork's painterly style and generate texture with similar brush strokes. However, directly transferring the style of an entire painting can lead to texture mismatching problem since the semantic context of the foreground and background should be considered. Our key contributions include:
	
	\begin{itemize}
		\item A \emph{painting estimator} to estimate the level of realism of a given painting by fine-tuning pre-trained VGG-16~\cite{simonyan2014very} on $\sim 80$k paintings collected from \texttt{wikiart.org};
		\item A \emph{painting optimizer} which first selects texture source patches of a painting by computing the correspondence field semantically in deep neural space and then synthesizes the compositing result through optimization.
	\end{itemize}
	
	We demonstrate the practical impact of the proposed method on a large set of paintings with various styles including abstract art, baroque, cubism, expressionism, high renaissance, post-impressionism, realism, etc. and show that our method outperforms the state-of-the-arts. 
}


  \subsection{Related Work}

\paragraph{Image Harmonization.} The simplest way to blend images is
to combine the foreground and background color values using linear
interpolation, which is often accomplished using alpha
matting~\cite{porter1984compositing}. Gradient-domain compositing (or
Poisson blending) was first introduced by P{\'e}rez et
al.~\shortcite{perez2003poisson} which considers the boundary
condition for seamless cloning. Xue et
al.~\shortcite{xue2012understanding} identified key statistical
factors that affect the realism of photo compositings such as
luminance, color temperature, saturation, and local contrast, and
matched the histograms accordingly. Deep neural networks~\cite{zhu2015learning,tsai2017deep} further improved color properties of the composite by learning to improve the overall photo realism. Multi-Scale Image
Harmonization~\cite{sunkavalli2010multi} introduced smooth histogram
and noise matching which handles fine texture on top of color,
however it does not capture more structured textures like brush
strokes which often appear in paintings. Image Melding~\cite{darabi2012image} combines  Poisson
blending with patch-based synthesis~\cite{barnes2009patchmatch} in a
unified optimization framework to harmonize color and patch
similarity. Camouflage Images~\cite{chu2010camouflage} proposed an
algorithm to embed objects into certain locations in cluttered
photographs with a goal to make the objects hard to notice. 
While these techniques are mostly designed with photographs in mind, \fujun{our
focus is on paintings. In particular, we are interested in the case
where the background of the composite is a painting or a drawing.}

\paragraph{Style Transfer using Neural Networks.} Recent work on Neural Style
transfer~\cite{gatys2015neural} has shown impressive results on transferring
the style of an artwork by matching the statistics of layer responses of a deep
neural network. \fujun{These methods transfer arbitrary styles from one image to another by matching the correlations between feature activations extracted by a pretrained deep neural network on image classification (i.e., VGG~\cite{simonyan2014very}). The reconstruction process is based on an iterative optimization framework that minimizes the content and style losses computed from the VGG neural network. Recently, feed-forward generators propose fast approximations of the original Neural Style formulations~\cite{ulyanov2016texture,johnson2016perceptual,li2016precomputed} to achieve real-time performance. }
However, this technique is sensitive to mismatches in the image
content and several approaches have been proposed to address this
issue. Gatys et al.~\shortcite{Gatys2017a} add
the possibility for users to guide the transfer with annotations. In
the context of photographic transfer, Luan et
al.~\shortcite{luan2017deep} limit mismatches using scene
analysis. \fujun{Li and Wand~\shortcite{li2016combining} use 
nearest-neighbor correspondences between neural responses to make the
transfer content-aware. Specifically, they use a non-parametric
model that independently matches the local patches in each layer of
the neural network using normalized cross-correlation. Note that this
differs from our approach since we use feature
representations based on Gram matrices and enforce spatial consistency across different layers in the neural network when computing the correspondence. Improvements can be seen in the comparison results in Section~\ref{sec:first-pass}.} 
Odena et al.~\shortcite{odena2016deconvolution} study the filters used
in these networks and explain how to avoid the grid-like
artifacts produced by some techniques.
Recent approaches replace the Gram matrix with matching other statistics of neural responses~\cite{huang2017adain,Yijun2017Universal}.
Liao et al.~\shortcite{liao2017visual} further improve the
quality of the results by introducing bidirectional dense
correspondence field matching. 
%
All these methods have in common that they change the style of entire images
at once. Our work differs in that we focus on local transfer; we
shall see that global methods do not work as well when applied locally.


  \subsection{Background}

Our work builds upon the style transfer technique introduced by Gatys
et al.~\shortcite{gatys2015neural} (Neural Style) and several additional reconstruction losses proposed
later to improve its results. We summarize these techniques below
before describing our algorithm in the next section~(\S~\ref{sec:paint-harm-algor}).

\subsubsection{Style Transfer}

\fujun{Parts of our technique have a similar structure to the Neural
  Style algorithm by Gatys et al.~\cite{gatys2015neural}. They found
  that recent deep neural networks can learn to extract high-level semantic
  information and are able to independently manipulate the content and
  style of natural images. For completeness, we summarize the Neural
  Style algorithm that transfers a \emph{style} image to an
  \emph{input} image to produce an output image by minimizing loss
  functions defined using the VGG network. The algorithm proceeds in three
steps:} 
\begin{enumerate}
\itemsep-0.75\baselineskip
\item The input image $\img$ and style $\style$ are processed with the
  VGG network~\cite{simonyan2014very} to produce a set of activation
  values \fujun{as feature representations} $\neuralcoef[\img]$ and $\neuralcoef[\style]$. Intuitively,
  these capture the statistics that represent the style of each image.\\
\item The style activations are mapped to the input ones. In the
  original approach by Gatys et al., the entire set of style
  activations is used. Other options have been later proposed, e.g.,
  using nearest neighbors neural patches~\cite{li2016combining}.\\
\item The output image $\out$ is reconstructed through an optimization
  process that seeks to preserve the content of the input image while
  at the same time match the visual appearance of the style
  image. These objectives are modeled using \emph{losses} that we
  describe in more detail in the next section.\\
\end{enumerate}
\vspace{-\baselineskip} 
Our approach applies this three-step process twice,
the main variation being the activation matching step (2). Our first
pass uses a matching algorithm designed for robustness to large style
differences, and our second pass uses a more constrained matching designed to
achieve high visual quality.

\subsubsection{Reconstruction Losses}
\label{sec:reconstr-loss}

The last step of the pipeline proposed by Gatys et
al. is the reconstruction of the final
image $\out$. As previously discussed, this involves solving an
optimization problem that balances several objectives, each of them
modeled by a \emph{loss function}. Originally, Gatys et al.\ proposed
two losses: one to preserve the content of the input image $\img$ and one to
match the visual appearance of the style image~$\style$. Later, more
reconstruction losses have been proposed to improve the quality of the
output. Our work builds upon several of them that we review below.

\paragraph{Style and Content Losses.}
\label{sec:style-content-losses}
In their original work, Gatys et al.\ used the loss below.
\begin{subequations}
\label{eq:LGatys}
\begin{align}
\label{eq:neural_style}
\Gatysloss &= 
\contentloss + \styleweight\styleloss
\\[10pt]
\label{eq:neural_style_c} \text{with: } \contentloss & =
                                                       \sum_{\neurallayer=1}^{\numberoflayers} 
\frac{\alpha_\neurallayer}{2\numberoffilters_\neurallayer\numberofcoefs_\neurallayer}
                                           \sum_{\neuralfilter=1}^{\numberoffilters_\neurallayer}
                                           \sum_{\neuralpixel=1}^{\numberofcoefs_\neurallayer}\big(\neuralcoef_\neurallayer[\out] - \neuralcoef_\neurallayer[\img]\big)_{\neuralfilter\neuralpixel}^2
\\
\label{eq:neural_style_s}
\styleloss & = \sum_{\neurallayer=1}^{\numberoflayers}
              \frac{\beta_\neurallayer}{2\numberoffilters_\neurallayer^2} \sum_{\neuralfilter=1}^{\numberoffilters_\neurallayer}\sum_{\neuralfilterp=1}^{\numberoffilters_\neurallayer}\big(\gram_\neurallayer[\out] - \gram_\neurallayer[\style]\big)_{\neuralfilter\neuralfilterp}^2
\end{align}
\end{subequations}
where $\numberoflayers$ is the total number of convolutional layers,
$\numberoffilters_\neurallayer$ the number of filters in the
$\neurallayer^\text{th}$ layer, and $\numberofcoefs_\neurallayer$ the
number of activation values in the filters of the
$\neurallayer^\text{th}$
layer. $\neuralcoef_\neurallayer[\cdot]\in\mathbb{R}^{\numberoffilters_\neurallayer\times\numberofcoefs_\neurallayer}$
is a matrix where the $(\neuralfilter,\neuralpixel)$ coefficient is the $\neuralpixel^\text{th}$ activation of the
$\neuralfilter^\text{th}$ filter of the $\neurallayer^\text{th}$ layer
and
$\gram_\neurallayer[\cdot]=\neuralcoef_\neurallayer[\cdot]\transpose{\neuralcoef_\neurallayer[\cdot]}
\in
\mathbb{R}^{\numberoffilters_\neurallayer\times\numberoffilters_\neurallayer}$
is the corresponding Gram matrix. 
$\alpha_\neurallayer$ and $\beta_\neurallayer$ are weights controlling
the influence of each layer and $\styleweight$ controls the tradeoff
between the {\em content} (Eq.~\ref{eq:neural_style_c}) and the {\em
  style} (Eq.~\ref{eq:neural_style_s}). The advantage of the Gram
matrices $\gram_\neurallayer$ is that they represent the statistics of
the activation values $\neuralcoef_\neurallayer$ independently of
their location in the image, thereby allowing the style statistics to be
``redistributed'' in the image as needed to fit the input
content. Said differently, the product
$\neuralcoef_\neurallayer[\cdot]\transpose{\neuralcoef_\neurallayer[\cdot]}$
  amounts to summing over the entire image, thereby pooling local
  statistics into a global representation.

\paragraph{Histogram Loss.} Wilmot et al.~\shortcite{wilmot2017stable}
showed that $\Gatysloss$ is unstable because of ambiguities inherent
in the Gram matrices and proposed the loss below to ensure that 
activation histograms are preserved, which remedies the ambiguity.

\begin{subequations}
\begin{align}
\label{eq:histogram_loss}
\histloss &=  \sum_{\neurallayer=1}^{\numberoflayers} \gamma_\neurallayer \sum_{\neuralfilter=1}^{\numberoffilters_\neurallayer}
                                           \sum_{\neuralpixel=1}^{\numberofcoefs_\neurallayer} \big(\neuralcoef_\neurallayer[\out] - \histmatch_\neurallayer[\out]\big)_{\neuralfilter\neuralpixel}^2
\\[10pt]
\text{with: }
\histmatch_\ell[\out] &= \texttt{histmatch}(\neuralcoef_\neurallayer[\out], \neuralcoef_\neurallayer[\style])
\end{align}
\end{subequations}
where $\gamma_\ell$ are  weights controlling the influence of each layer and $\histmatch_\neurallayer[\out]$ is the histogram-remapped feature map by matching $\neuralcoef_\neurallayer[\out]$ to $\neuralcoef_\neurallayer[\style]$.

\paragraph{Total Variation Loss.} Johnson et
al.~\shortcite{johnson2016perceptual} showed that the total variation loss
introduced by Mahendran and
Vedaldi~\shortcite{mahendran2015understanding} improves style
transfer results by producing smoother outputs.
\begin{equation}
\tvloss(\out) =  \sum_{x,y} (\out_{x,y}-\out_{x,y-1})^2 + (\out_{x,y}-\out_{x-1,y})^2
\label{eq:tv_loss}
\end{equation}
where the sum is over all the $(x,y)$ pixels of the output image $\out$.


  \section{Painterly Harmonization Algorithm}
\label{sec:paint-harm-algor}

We designed a two-pass algorithm to achieve painterly harmonization.
Previous work used a single-pass approach; for example, Gatys et
al.~\shortcite{gatys2015neural} match the entire style image to the entire
input image and then use the $L_2$ norm on Gram matrices to reconstruct the
final result. Li and Wand~\shortcite{li2016combining} use nearest neighbors for
matching and the $L_2$ norm on the activation vectors for reconstructing.  In
our early experiments, we found that such single-pass strategies did not work
as well in our context and we were not able to achieve as good results as we
hoped.
This motivated us to develop a two-pass approach where the first pass
aims for coarse harmonization, and the second focuses on fine visual quality
(Alg.~\ref{alg:entirepip}).  

The first pass 
produces an intermediate result that is
close to the desired style but we do not seek to produce the highest quality
output possible at this point. By relaxing the requirement of high quality, we
are able to design a robust algorithm that can cope with vastly different
styles.  This pass achieves coarse harmonization by first performing a
rough match of the color and texture properties of the pasted region to those
of semantically similar regions in the painting. We find nearest-neighbor
neural patches independently on each network layer (Alg.~\ref{alg:inconsistentmapping}) to
match the responses of the pasted region and of the background. This gives us
an intermediate result (Fig.~\ref{fig:stepone}b) that is a better starting point for
the second pass.

Then, in the second pass, we start from this intermediate result and focus on
visual quality. Intuitively, since the intermediate image and the style image
are visually close, we can impose more stringent requirements on the output
quality.  In this pass, we work at a single intermediate layer that captures the
local texture properties of the image. This generates a correspondence map that
we process to remove spatial outliers. We then upsample this spatially
consistent map to the finer levels of the network, thereby ensuring that at
each output location, the neural responses at all scales come from the same
location in the painting (Alg.~\ref{alg:consistentmapping}). This leads to more coherent
textures and better looking results (Fig.~\ref{fig:stepone}c).
In the rest of this section, we describe in detail each step
of the two passes.

\ignorethis{
\kb{We need to clean up all the ??? with the correct algorithm. I added a
placeholder for a 2pass version to make clear which output gets generated by
whom and what feeds to the next level, etc. We should talk through this. We
should also explain what pi is when it invokes this new root algorithm.}
}



\begin{algorithm}[htp]
  \caption{\textit{TwoPassHarmonization}$(\img,\mask,\style)$\label{alg:entirepip}}
  \DontPrintSemicolon
  \SetKw{KwData}{input}
  \SetKw{KwResult}{output}
  \SetFuncSty{textit}
  \SetCommentSty{textrm}
  \SetKwFunction{applyVGG}{ComputeNeuralActivations}
  \SetKwFunction{mapcoefs}{\neuralmapping}
  \SetKwFunction{reconstruct}{Reconstruct}
  \SetKwFunction{singlePassHarmonization}{SinglePassHarmonization}
  \SetKwFunction{independentMapping}{IndependentMapping}
  \SetKwFunction{consistentMapping}{ConsistentMapping}
  
\begin{tabular}{ll}
    \KwData & input image $\img$ and mask $\mask$ \\
            & style image $\style$ \\
    \KwResult {\Large \strut} & output image $\out$
  \end{tabular}
  \BlankLine


  \linecomment{Pass \#1: Robust coarse harmonization
    (\S~\ref{sec:first-pass}, Alg.~\ref{alg:single_pass}) \\ Treat each layer
  independently during input-to-style mapping (Alg.~\ref{alg:inconsistentmapping})}
$\inter \leftarrow \singlePassHarmonization(\img, \mask, \style, \independentMapping)$ \;
  \BlankLine 
\linecomment{Pass \#2: High-quality refinement
    (\S~\ref{sec:second-pass}, Alg.~\ref{alg:single_pass}) \\ Enforce
    consistency across layers \\ and in image space during
    input-to-style mapping (Alg.~\ref{alg:consistentmapping})}
  $\out \leftarrow \textit{SinglePassHarmonization}(\inter, \mask, \style, \consistentMapping)$ \;
\end{algorithm}


\begin{algorithm}[t]
  \caption{\textit{SinglePassHarmonization}$(\img,\mask,\style,\neuralmapping)$\label{alg:single_pass}} 
  \DontPrintSemicolon
  \SetKw{KwData}{input}
  \SetKw{KwResult}{output}
  \SetFuncSty{textit}
  \SetCommentSty{textrm}
  \SetKwFunction{applyVGG}{ComputeNeuralActivations}
  \SetKwFunction{mapcoefs}{\neuralmapping}
  \SetKwFunction{reconstruct}{Reconstruct}
  
\begin{tabular}{ll}
    \KwData & input image $\img$ and mask $\mask$ \\
            & style image $\style$ \\
            & neural mapping function $\mapcoefs$ \\
    \KwResult {\Large \strut} & output image $\out$
  \end{tabular}
  \BlankLine
  \linecomment{Process input and style images with VGG network.}
  $\neuralcoef[\img] \leftarrow \applyVGG(\img)$ \;
  $\neuralcoef[\style] \leftarrow \applyVGG(\style)$ \;
  \BlankLine
  \linecomment{Match each input activation in the mask to a style
    activation\\ and store the mapping from the former to the latter in $\stylemap$.}
  $\stylemap \leftarrow
  \mapcoefs(\neuralcoef[\img],\mask,\neuralcoef[\style])$ \;
  \BlankLine
  \linecomment{Reconstruct output image to approximate new
    activations.}
  $\out \leftarrow \reconstruct(\img,\mask,\style,\stylemap)$ \;
\end{algorithm}



\subsection{First Pass: Robust Coarse Harmonization}
\label{sec:first-pass}

We designed our first pass to be robust to the diversity of paintings
that users may provide as style images. In our early experiments, we
made two observations. First, we applied the technique of Gatys et
al.~\shortcite{gatys2015neural} as is, that is, we used the entire
style image to build the style loss $\styleloss$. This produced
results where the pasted element became a ``summary'' of the style
image. For instance, with Van Gogh's Starry Night, the pasted element
had a bit of swirly sky, one shiny star, some of the village
structure, and a small part of the wavy trees. While each texture was
properly represented, the result was not satisfying because only a
subset of them made sense for the pasted element. Then, we
experimented with the nearest-neighbor approach of Li and
Wand~\shortcite{li2016combining}. The intuition is that by assigning
the closest style patch to each input patch, it selects style
statistics more relevant to the pasted element. Although the generated
texture tended to lack contrast compared to the original painting, the
results were more satisfying because the texture was more
appropriate. Based on these observations, we designed the algorithm
below that relies on nearest-neighbor correspondences and a
reconstruction loss adapted from~\cite{gatys2015neural}.

 \begin{figure}[htp]
    \centering
    \begin{subfigure}{.162\textwidth}
      \includegraphics[width=0.95\linewidth]{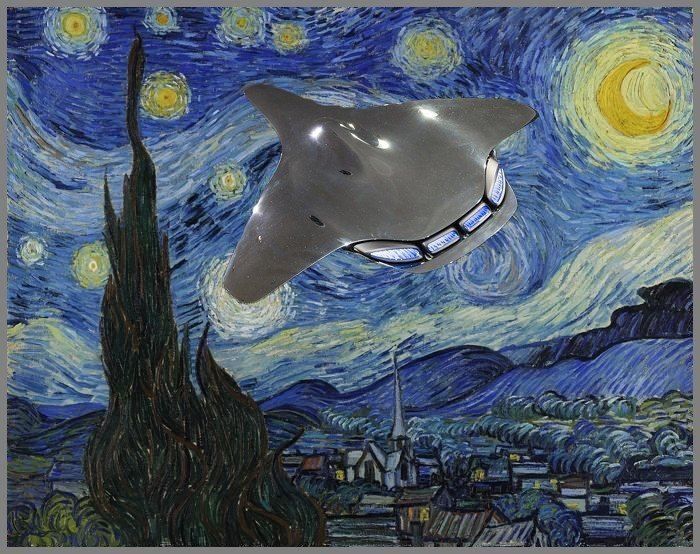}
    \end{subfigure}%
    \begin{subfigure}{.162\textwidth}
      \includegraphics[width=0.95\linewidth]{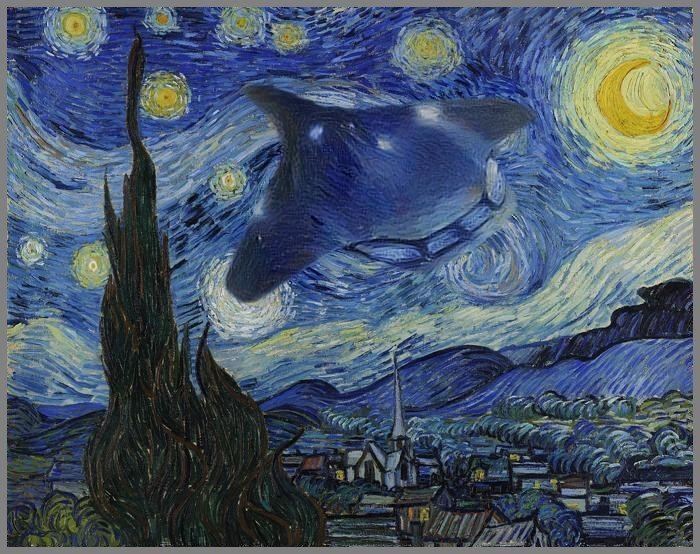}
    \end{subfigure}%
    \begin{subfigure}{.162\textwidth}
      \includegraphics[width=0.95\linewidth]{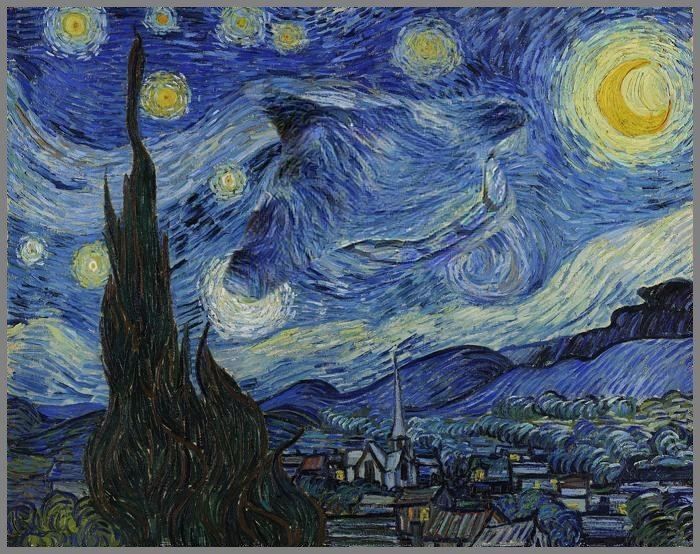}
    \end{subfigure}\vspace{1mm}

    \begin{subfigure}[t]{.162\textwidth}
      \includegraphics[width=0.95\linewidth]{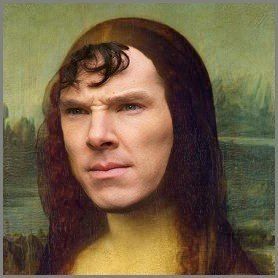}
      \caption{\raggedright Cut-and-paste}
    \end{subfigure}%
    \begin{subfigure}[t]{.162\textwidth}
      \includegraphics[width=0.95\linewidth]{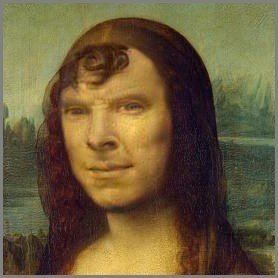}
      \caption{\raggedright $1^\text{st}$ pass.  Robust harmonization
        but \newline weak texture  (top) \newline and artifacts (bottom).}
    \end{subfigure}%
    \begin{subfigure}[t]{.162\textwidth}
      \includegraphics[width=0.95\linewidth]{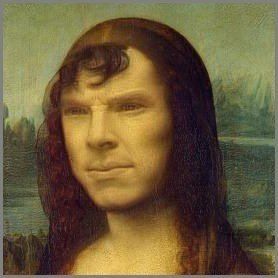}
      \caption{\raggedright $2^\text{nd}$ pass. Refined results with accurate
        texture and no artifact.}
    \end{subfigure}\vspace{1mm}
  
    \caption{Starting from vastly different input and style
      images~(a), we first harmonize the overall appearance of the
      pasted element~(b) and then refine the result to finely match
      the texture and remove artifacts~(c). }
    \label{fig:stepone}
 
  \end{figure}

\paragraph{Mapping.}
\label{sec:first-pass-mapping}
Similarly to Li and Wand, for each layer $\neurallayer$ of the neural
network, we stack the activation coefficients at the same location in
the different feature maps into an \emph{activation vector}. Instead
of considering $\numberoffilters_\neurallayer$ feature maps, each of
them with $\numberofpatches_\neurallayer$ coefficients, we work with a
single map that contains $\numberofpatches_\neurallayer$ activation
vectors of dimension~$\numberoffilters_\neurallayer$. For each
activation vector, we consider the $3\times3$ patch centered on it. We
use nearest neighbors based on the $L_2$ norm on these patches to assign
a style vector to each input vector. We call this strategy
\emph{independent mapping} because the assignment is made
independently for each layer. Algorithm~\ref{alg:inconsistentmapping}
gives the pseudocode of this mapping. Intuitively, the independence
across layers makes the process more robust because a poor match in a
layer can be compensated for by better matches in the other
layers. The downside of this approach is that the lack of coherence
across layers impacts the quality of the output
(Fig.~\ref{fig:stepone}b). However, as we shall see, these
artifacts are limited and our second pass removes them.

\begin{figure}[htp]
\centering
\newcommand{\imgH}{2.75cm}
      \begin{subfigure}[t]{.5\columnwidth}
        \includegraphics[height=\imgH]{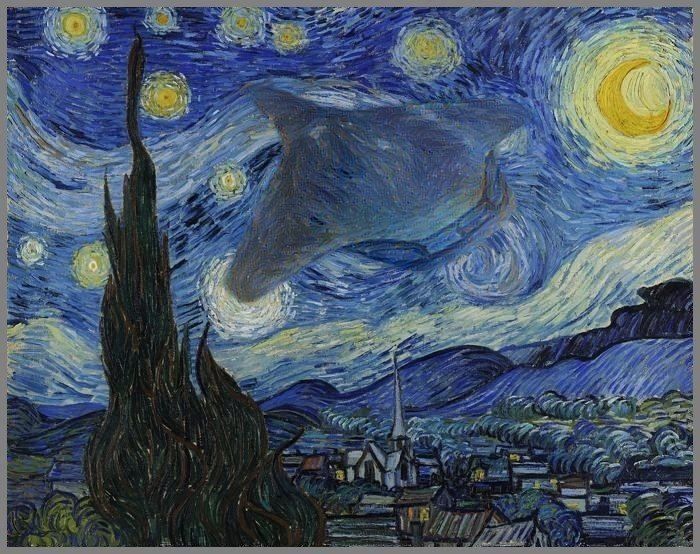}
        \caption*{\hspace{-6ex} Overly weak texture}
      \end{subfigure}%
      \begin{subfigure}[t]{.35\columnwidth}
        \includegraphics[height=\imgH]{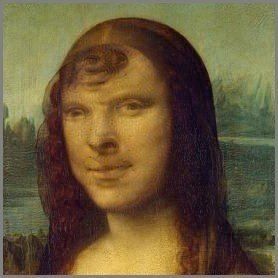}
        \caption*{\hspace{-3ex} Severe artifacts}
      \end{subfigure}%
      \caption{Examples of quality loss when not using a Gram matrix
        in the first pass. The inputs are the same as in
        Figure~\ref{fig:stepone}. \fujun{Directly applying the $L_2$ norm,
          similarly to Li and Wand, forces each spatial location to
          use only the nearest neighbor from the style image, which
          causes artifacts when the match is not good. As we show later, using a Gram matrix to pack the matched features together relaxes this limitation and produces fewer artifacts.}}
      \label{fig:L2_recons}
\end{figure}

\paragraph{Reconstruction. } \label{sec:first-pass-reconstruction}
Unlike Li and Wand who use the $L_2$ norm on these activation vectors
to reconstruct the output image, we pool the vectors into Gram
matrices and use $\Gatysloss$ (Eq.~\ref{eq:LGatys}). Applying the
$L_2$ norm directly on the vectors constrains the spatial
location of the activation values; the Gram matrices relax this
constraint as discussed in \S~\ref{sec:style-content-losses}. Figure~\ref{fig:L2_recons} shows
that using $L_2$ reconstruction directly, i.e.,
without Gram matrices, does not produce as good~results.

\begin{algorithm}
  \caption{\textit{IndependentMapping}$(\neuralcoef[\img],\mask,\neuralcoef[\style])$
  \label{alg:inconsistentmapping}}
  \DontPrintSemicolon
  \SetKw{KwData}{input}
  \SetKw{KwResult}{output}
  \SetFuncSty{textit}
  \SetCommentSty{textrm}
  \SetKwFunction{nnindex}{NearestNeighborIndex}
  \SetKwFunction{resize}{Resize}
  
\begin{tabular}{ll}
    \KwData & input neural activations $\neuralcoef[\img]$ and mask $\mask$ \\
            & style neural activations $\neuralcoef[\style]$ \\
    \KwResult {\Large \strut} & input-to-style mapping $\stylemap$
  \end{tabular}
  \BlankLine
  \linecomment{For each layer in the network...}
  \For(\eolcomment{$\numberoflayers=\text{number of layers}$}){$\neurallayer\in[1:\numberoflayers]$}{
    \BlankLine
    \linecomment{For each ``activation patch'' in the
      $\neurallayer^\text{th}$ layer...}
    \linecomment{\hspace{3ex} \it ``activation patch'' = vector made of all
      the activations
     \\ \hspace{4ex} in a
      $ 3\times3$ patch across all the filters of a layer.}
    \For(\eolcomment{$\numberofpatches_\neurallayer=$ number of
        patches in the $\neurallayer^\text{th}$ layer}){$\neuralpatch\in[1:\numberofpatches_\neurallayer]$}{
      \BlankLine  
      \linecomment{Consider only the patches inside the mask \\
          resized to the resolution of the $\neurallayer^\text{th}$ layer}
        \If{$\neuralpatch\in\resize(\mask,\neurallayer)$}{
          \BlankLine
          \linecomment{Assign the style patch closest to the input patch}
          $\stylemap(\neurallayer,\neuralpatch) \leftarrow
          \nnindex(\neuralcoef_\neurallayer[\img]_{\neuralpatch},
          \neuralcoef_\neurallayer[\style])$ \; 
        } 
      } 
    }
\end{algorithm}

\subsection{Second Pass: High-Quality Refinement}
\label{sec:second-pass}
\begin{figure*}[htp]
   \centering \subcaptionbox{Style image}[.23\textwidth]{\includegraphics[width=0.23\linewidth]{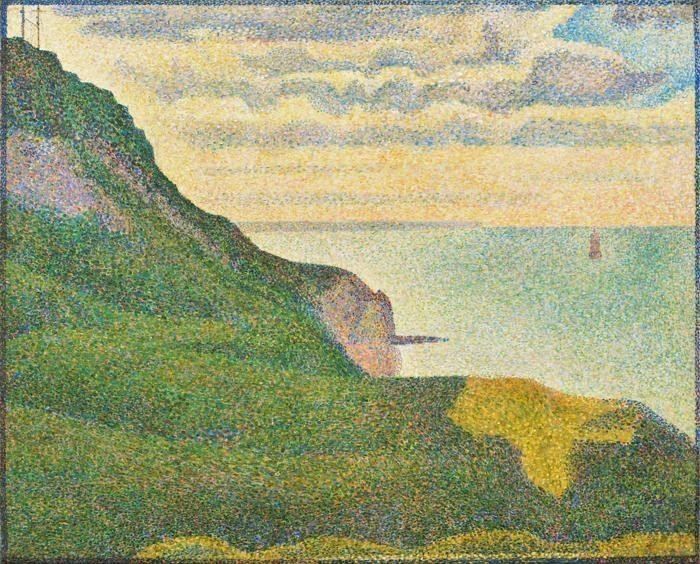}\vspace{-0.25\baselineskip}}\hspace{0.2cm}
\subcaptionbox{Cut-and-paste}[.23\textwidth]{\includegraphics[width=0.23\linewidth]{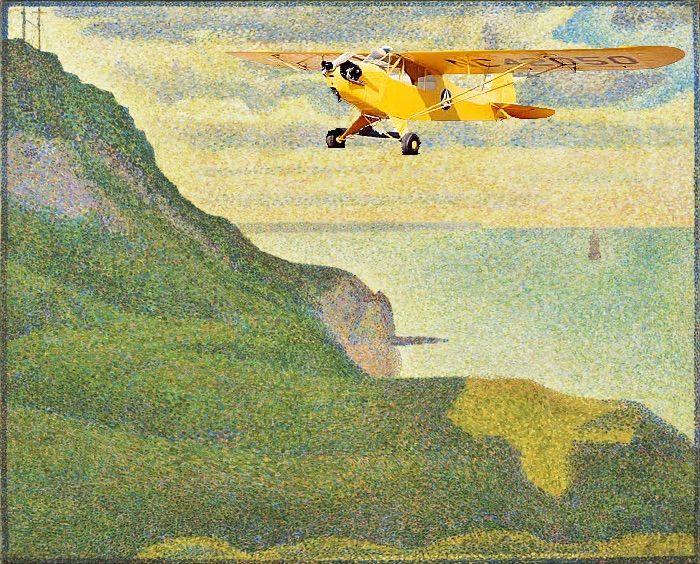}\vspace{-0.25\baselineskip}}\hspace{0.2cm}
\subcaptionbox{Independent mapping ($1^\text{st}$ pass only, our intermediate result)} [.23\textwidth]{\includegraphics[width=0.23\linewidth]{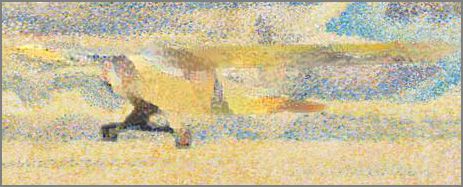}\vspace{-0.25\baselineskip}}\hspace{0.2cm}
\subcaptionbox{Consistent mapping ($2^\text{nd}$ pass only, bad correspondence)}[.23\textwidth]{\includegraphics[width=0.23\linewidth]{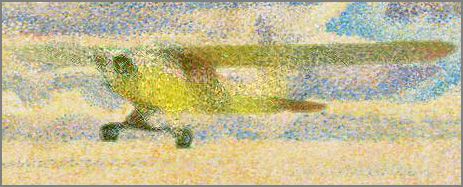}\vspace{-0.25\baselineskip}}
\\[0.5\baselineskip]
\subcaptionbox{Entire pipeline without $\histloss$ and using $\styleloss$ instead of $\uniquestyleloss$}[.23\textwidth]{\includegraphics[width=0.23\linewidth]{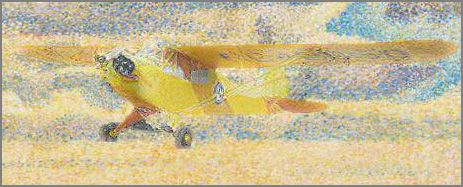}\vspace{-0.25\baselineskip}}\hspace{0.2cm}
\subcaptionbox{Entire pipeline using $\styleloss$ \newline instead of
  $\uniquestyleloss$}[.23\textwidth]{\includegraphics[width=0.23\linewidth]{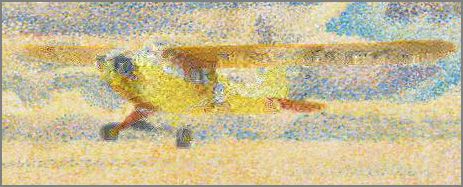}\vspace{-0.25\baselineskip}}\hspace{0.2cm}
\subcaptionbox{Entire pipeline without painting estimator (default parameters, style is too weak)}[.23\textwidth]{\includegraphics[width=0.23\linewidth]{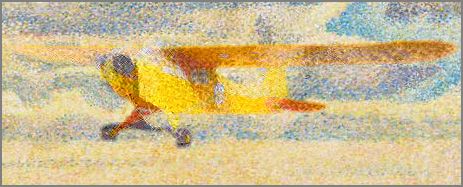}\vspace{-0.25\baselineskip}}\hspace{0.2cm}
\subcaptionbox{Our final result}[.23\textwidth]{\includegraphics[width=0.23\linewidth]{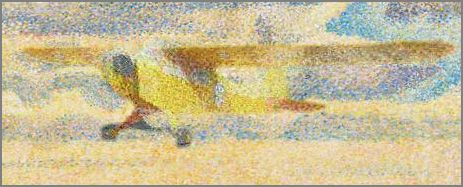}\vspace{-0.25\baselineskip}}
\caption{Ablation study. (c-h) are cropped for best fit. Zoom in for
  details. \fujun{The $1^\text{st}$ pass (a) reduces the style
    difference gap between the foreground and background, but lacks fine
    texture details. Directly applying the $2^\text{nd}$ pass produces
    fine texture details, but the correspondence is not as good, which
    causes texture mismatch problems. Without the histogram loss, (e)
    is unable to reproduce the dot texture of the background painting
    in some regions due to the instabilities of the Gram
    matrix. Without addressing many-to-one mappings, redundant style
    patches are reused multiple times and cause artifacts in the
    output (f). Without the painting estimator, the style of the result
    (g) is not well reproduced. (h) is our final solution by combining all components. }}
\label{fig:ablation_study}
\end{figure*}

As can be seen in Figure~\ref{fig:stepone}(b), the results after the
first pass match the desired style but suffer from artifacts. In our
early experiment, we tried to fine-tune the first pass but our attempts
only improved some results at the expense of others. Adding
constraints to achieve a better quality was making the process less robust to
style diversity. We address this challenge with a second pass that
focuses on visual quality. The advantage of starting a complete new
pass is that we now start from an intermediate image close to the
desired result and robustness is not an issue anymore. 
We design our pass such that the input-to-style mapping is consistent across
layers and space. We ensure that the activation vectors assigned to
the same image location on different layers were already collocated in
the style image. We also favor the configuration where vectors adjacent in the style image remain adjacent in the
mapping. Enforcing such strict requirements directly on the input
image often yields poor results (Fig.~\ref{fig:ablation_study}d) but
when starting from the intermediate image generated by the first pass,
this approach produces high quality outputs
(Fig.~\ref{fig:ablation_study}h). We also build on previous work to
improve the reconstruction step. We explain the details of each step below.

\begin{figure}[htp]
    \centering

    \begin{subfigure}{.162\textwidth}
      \includegraphics[width=0.95\linewidth]{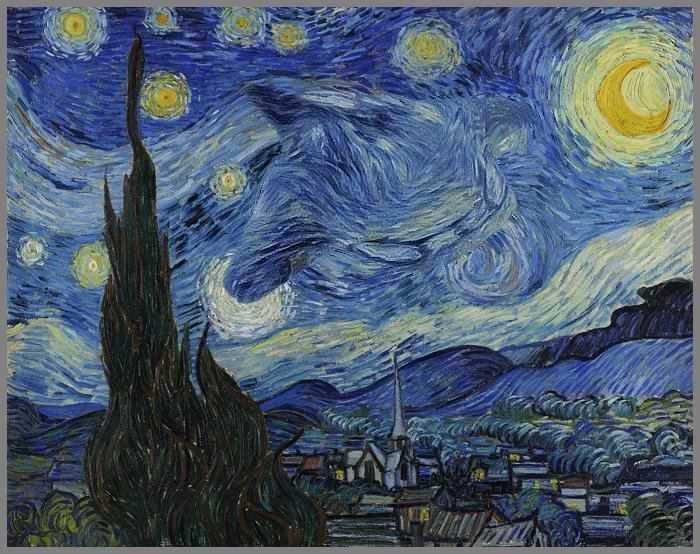}
    \end{subfigure}%
    \begin{subfigure}{.162\textwidth}
      \includegraphics[width=0.95\linewidth]{free2use/2_result_conv4_1}
    \end{subfigure}%
    \begin{subfigure}{.162\textwidth}
      \includegraphics[width=0.95\linewidth]{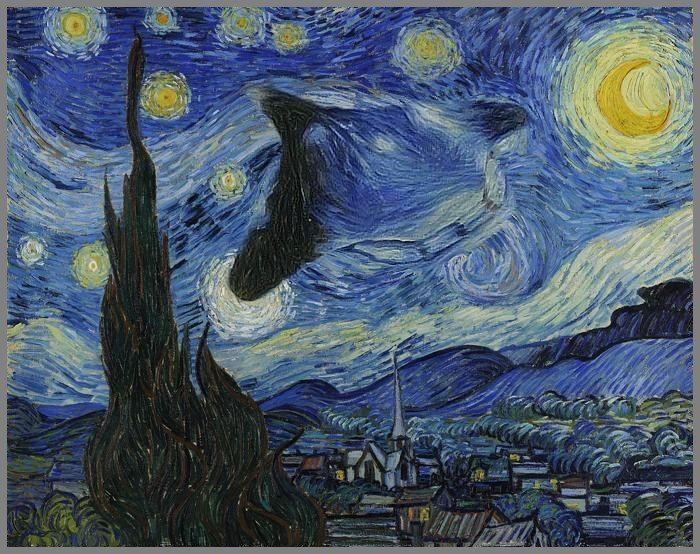}
    \end{subfigure}\vspace{1mm}

    \begin{subfigure}[t]{.162\textwidth}
      \includegraphics[width=0.95\linewidth]{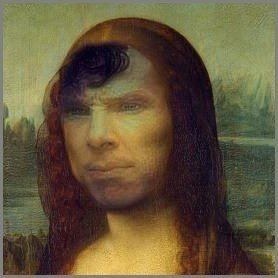}
      \caption{$\refneurallayer = \vgglayer{conv3\_1}$}
    \end{subfigure}%
    \begin{subfigure}[t]{.162\textwidth}
      \includegraphics[width=0.95\linewidth]{free2use/1_result2_crop}
      \caption{\centering$\refneurallayer = \vgglayer{conv4\_1}$\newline (our setting)}
    \end{subfigure}%
    \begin{subfigure}[t]{.162\textwidth}
      \includegraphics[width=0.95\linewidth]{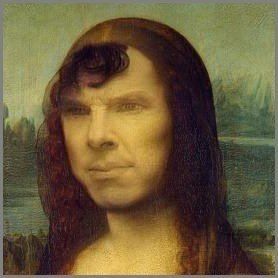}
      \caption{$\refneurallayer = \vgglayer{conv5\_1}$}
    \end{subfigure}\vspace{1mm}
 
    \caption{Setting $\refneurallayer$ to \vgglayer{conv3\_1} produces low-quality results due to poor matches between the input and style images~(a). Instead we use \vgglayer{conv4\_1} that yields better results~(b). Using the deeper layer \vgglayer{conv5\_1} generates lower-quality texture~(c) but the degradation is minor compared to using \vgglayer{conv3\_1}. The inputs are the same as in Figure~\ref{fig:stepone}.}
    \label{fig:ref_layer}
  \end{figure}

\paragraph{Mapping.}
We start with a nearest-neighbor assignment similar to the first pass
(\S~\ref{sec:first-pass-mapping}) but applied only to a single layer,
which we call the \emph{reference layer} $\refneurallayer$
(Alg.~\ref{alg:consistentmapping}, Step~\#1).  We
tried several layers for $\refneurallayer$ and found that
\vgglayer{conv4\_1} provided a good trade-off between deeper layers
that ignore texture, and shallower layers that ignore scene semantics,
e.g., pairing unrelated objects together (Fig.~\ref{fig:ref_layer}).

\begin{algorithm}[htp]
  \caption{\textit{ConsistentMapping}$(\neuralcoef[\img],\mask,\neuralcoef[\style])$\label{alg:consistentmapping}}
  \DontPrintSemicolon
  \SetKw{KwData}{input}
  \SetKw{KwResult}{output}
  \SetFuncSty{textit}
  \SetDataSty{textit}
  \SetCommentSty{textrm}
  \SetKwFunction{nn}{NearestNeighbor}
  \SetKwFunction{nnindex}{NearestNeighborIndex}
  \SetKwFunction{resize}{Resize}
  \SetKwFunction{changeres}{ChangeResolution}
  \SetKwData{candidateSet}{CSet}

\begin{tabular}{ll}
    \KwData & input neural coefficients $\neuralcoef[\img]$ and mask
              $\mask$ \\
            & style neural coefficients $\neuralcoef[\style]$ \\
    \KwResult {\Large \strut} & input-to-style mapping $\stylemapout$
  \end{tabular}
  \BlankLine
  \linecomment{Step \#1: Find matches for the reference layer.}
  \linecomment{Do the same as in
    Alg.~\ref{alg:inconsistentmapping} 
    but only for the reference layer. 
  }
  \For{$\neuralpatch\in[1:\numberofpatches_\refneurallayer]$}{
    \BlankLine
    \If{$\neuralpatch\in\resize(\mask,\refneurallayer)$}{
      \BlankLine
      \linecomment{$\stylemap$ is an intermediate input-to-style
        mapping refined \\ in the next step of the algorithm.}
      $\stylemap(\refneurallayer,\neuralpatch) \leftarrow
      \nnindex(\neuralcoef_\refneurallayer[\img]_{\neuralpatch},
      \neuralcoef_\refneurallayer[\style])$ \;
    } 
  }
  \BlankLine
  \linecomment{Step \#2: Enforce spatial consistency.}
  \For{$\neuralpatch\in[1:\numberofpatches_\refneurallayer]$}{
    \BlankLine  
    \If{$\neuralpatch\in\resize(\mask,\refneurallayer)$}{
      \BlankLine
      \linecomment{Look up the corresponding style patch.}
      $\neuralstylepatch \leftarrow \stylemap(\refneurallayer,\neuralpatch)$\;
      \BlankLine
      \linecomment{Initialize a set of candidate style patches.}
      $\candidateSet \leftarrow
      \{\neuralstylepatch\}$\;
      \BlankLine
      \linecomment{For all adjacent patches...}
      \For{$\offset\in\adjoffsetset$}{        
        \BlankLine
        \linecomment{Duplicate its assignment, i.e.:
          \\ 1. Look up the style patch of the adjacent patch
          $\neuralpatch+\offset$ 
          \\ \hspace{2ex} and apply the opposite offset $-\offset$.
          \\ 2. Add the result to the set of candidates.}
        $\candidateSet \leftarrow \candidateSet \cup
        \{\stylemap(\refneurallayer,\neuralpatch+\offset) - \offset\}$ \;
      }
      \BlankLine
      \linecomment{Select the candidate the most similar to the
        style patches 
        \\ associated to the neighbors of $\neuralpatch$.}
      $\displaystyle\stylemapout(\refneurallayer,\neuralpatch) \;\;
      \leftarrow \newline ~ \hfill \underset{\candidatepatch\in\candidateSet}{\argmin}\sum_{\offset\in\adjoffsetsetshort}
      \lVert\neuralcoef_\refneurallayer[\style]_\candidatepatch - 
      \neuralcoef_\refneurallayer[\style]_{\stylemap(\refneurallayer,\neuralpatch+\offset)}\rVert^2$\;
      \hfill\linecomment{with $\adjoffsetsetshort=\adjoffsetset$}
    }
  }
  \BlankLine
  \linecomment{Step \#3: Propagate the matches in the ref.\ layer
    to the other layers.}
  \linecomment{For each layer in the network excluding the reference layer...}
  \For{$\neurallayer\in[1:(\refneurallayer-1)]\cup[(\refneurallayer+1):\numberoflayers]$}{
    \BlankLine
    \For{$\neuralpatch\in[1:\numberofpatches_\neurallayer]$}{
      \BlankLine
      \If{$\neuralpatch\in\resize(\mask,\neurallayer)$}{
        \BlankLine
        \linecomment{Compute the index  of the
          patch in
            $\neuralcoef_\neurallayer[\img]$ \\  at the
            same image location as $\neuralcoef_\refneurallayer[\img]_\neuralpatch$.}
          $\refneuralpatch \leftarrow \changeres[\img](\neurallayer,\refneurallayer,\neuralpatch)$ \;
          \BlankLine
          \linecomment{Fetch matching style patch in the reference layer.}
          $\neuralstylepatch \leftarrow \stylemapout(\refneurallayer,\refneuralpatch)$ \;
          \BlankLine
          \linecomment{Change the resolution back.}
          $\stylemapout(\neurallayer,\neuralpatch)\leftarrow
          \changeres[\style](\refneurallayer,\neurallayer,\neuralstylepatch)$\;
        }
      }
    }

    \BlankLine
    
  \end{algorithm}


Then, we process this single-layer mapping to improve its spatial
consistency by removing outliers. We favor configurations where all
the style vectors assigned to an input region come from the same
region in the style image. For each input vector $\neuralpatch$, we
compare the style vector assigned by the nearest-neighbor
correspondence above as well as the vectors obtained by duplicating
the assignments of the neighbors of $\neuralpatch$. Among these
candidates, we pick the vector pointing to a style feature that is
most similar to its neighbors' features. In practice, this removes small
outlier regions that are inconsistent with their neighborhood. This procedure
is Step~\#2 of Algorithm~\ref{alg:consistentmapping}.

Last, we propagate these correspondences to the other layers so that
the activation values are consistent across layers. For a given
location in the input image, all the activation values across all the
layers are assigned style activations that come from the same location
in the style image (Alg.~\ref{alg:consistentmapping}, Step~\#3).

\paragraph{Reconstruction. }
A first option is to apply the same reconstruction as the first pass
(\S~\ref{sec:first-pass-reconstruction}), which already gives
satisfying results although some minor defects remain
(Fig.~\ref{fig:ablation_study}e). We modify the reconstruction as
follows to further improve the output. First, we observe that in some
cases, the nearest-neighbor assignment selects the same style vector
many times, which generates poor results. We address this issue by
selecting each vector at most once and ignoring the additional
occurrences, i.e., each vector contributes at most once to the Gram
matrix used in the style loss. We name $\uniquestyleloss$ this
variant of the style loss. We also add the histogram and
total-variation losses, $\histloss$ and $\tvloss$
(\S~\ref{sec:reconstr-loss}). Together, these form the loss that we
use to reconstruct our final output:
\begin{align}
  \finalloss &= \contentloss + \styleweight\uniquestyleloss +
               \histweight\histloss + \tvweight\tvloss
\end{align}
where the weights $\styleweight$, $\histweight$, and $\tvweight$
control the balance between the terms. Figure~\ref{fig:ablation_study} illustrates the
benefits of this loss. We explain in Section~\ref{sec:painting-estimator} how to set these
weights depending on the type of painting provided as the style image.

\paragraph{Discussion.}
Our constrained mapping was inspired by the nearest-neighbor field
upsampling used in the Deep Analogy work~\cite[\S~4.4]{liao2017visual}
that constrains the matches at each layer to come from a local region around the location at a previous layer. 
When the input and style images are similar, this technique
performs well. In our context, the intermediate image and the style
image are even more similar. This encouraged us to be even stricter by
forcing the matches to come from the exact same location. Beside this
similar aspect, the other algorithmic parts are different and as we shall
see, our approach produces better results for our application.

We experimented with using the same reconstruction in the first pass
as in the second pass. The quality gains were minimal on the
intermediate image and mostly non-existent on the final output. This
motivated us to use the simpler reconstruction in the first pass as described in
Section~\ref{sec:first-pass-reconstruction} for the sake of efficiency.

\begin{figure}[htp]
\centering

 \begin{subfigure}{.162\textwidth}
    \includegraphics[width=0.95\linewidth]{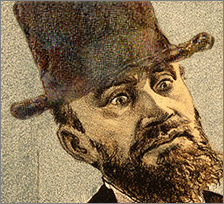}\vspace{-0.25\baselineskip}
    \caption{\fujun{Inset of our result \newline (before post-processing)}}
  \end{subfigure}%
  \begin{subfigure}{.162\textwidth}
    \includegraphics[width=0.95\linewidth]{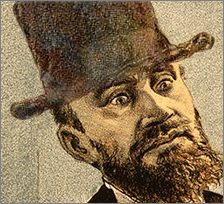}\vspace{-0.25\baselineskip}
    \caption{After guided filtering \newline{}}
  \end{subfigure}%
  \begin{subfigure}{.162\textwidth}
    \includegraphics[width=0.95\linewidth]{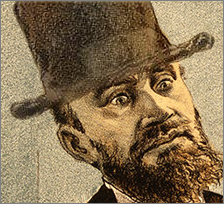}\vspace{-0.25\baselineskip}
    \caption{After patch synthesis \newline{}}
  \end{subfigure}
  \\[0.25\baselineskip]
  \begin{subfigure}[b]{.162\textwidth}
    \includegraphics[width=0.95\linewidth]{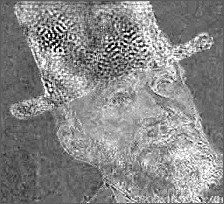}\vspace{-0.25\baselineskip}
    \caption{$a$ channel \newline (before guided filtering)}
  \end{subfigure}%
  \begin{subfigure}[b]{.162\textwidth}
    \includegraphics[width=0.95\linewidth]{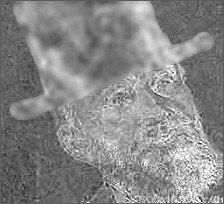}\vspace{-0.25\baselineskip}
    \caption{$a$ channel \newline (after guided filtering)}
  \end{subfigure}%
  \begin{subfigure}[b]{.162\textwidth}
    \includegraphics[width=0.95\linewidth]{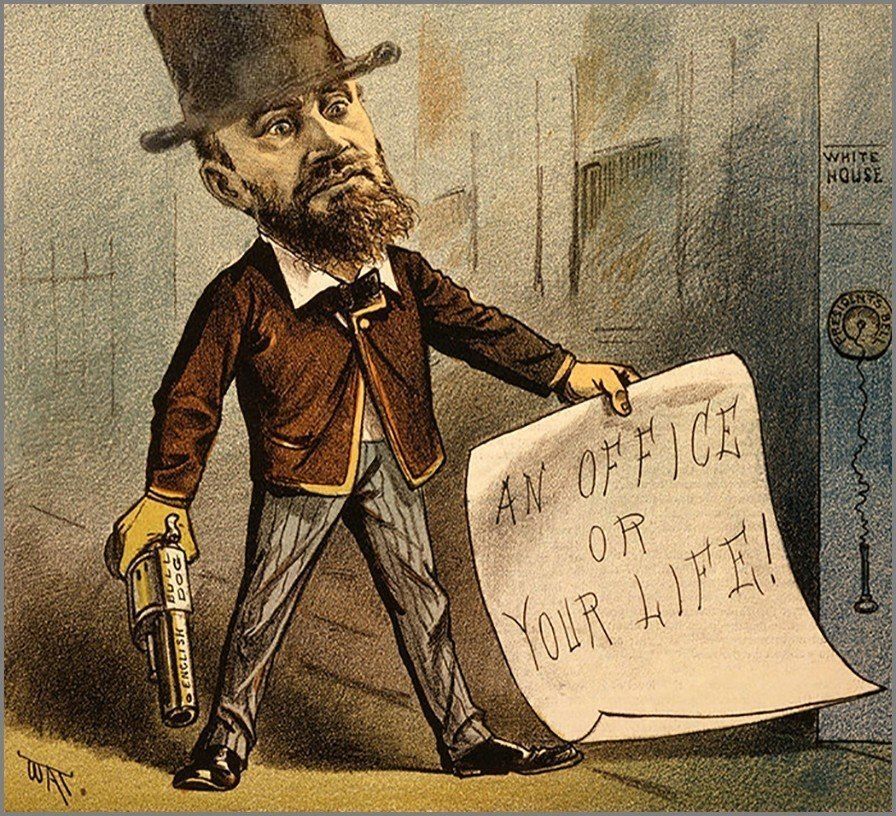}\vspace{-0.25\baselineskip}
    \caption{Our result \newline after post-processing}
  \end{subfigure}
 
\caption{Post-processing: Given deconvolution result with inset (a), we perform \emph{Chrominance Denoising} to produce (b) and \emph{Patch Synthesis} on (b) to produce (c). (d) and (e) show the insets of $a$ channel in CIE-Lab space before and after denosing. (f) is the final full-resolution result.}
\label{fig:cnn_artifact}
\end{figure}

\subsection{Post-processing}
  
The two-pass process described thus far yields high quality
results at medium and large scales but in some cases, fine-scale
details can be inaccurate. Said differently, the results are good from
a distance but may not be as satisfactory on close
examination.  The two-step signal-processing
approach below addresses this.

\paragraph{Chrominance Denoising. }
We observed that, in our context, high-frequency artifacts primarily affect
the chrominance channels while the luminance is
comparatively cleaner. We exploit this characteristic by converting
the image to CIE-Lab color space and applying the Guided
Filter~\cite{he2013guided} to filter the $ab$ chrominance channels
with the luminance channel as guide. We use the parameters suggested
by the authors, i.e., $r=2, eps=0.1^2$. This effectively suppresses the
highest-frequency color artifacts. However, some larger defects may
remain. The next step addresses this issue.

\paragraph{Patch Synthesis. }
The last step uses patch synthesis to ensure that every image patch in
the output appears in the painting. We use
PatchMatch~\cite{barnes2009patchmatch} to find a similar style patch
to each output patch. We reconstruct the output by averaging all overlapping
style patches, thereby ensuring that no new content is
introduced. However, the last averaging step tends to smooth
details. We mitigate this effect by separating the image into a
\emph{base layer} and a \emph{detail layer} using the Guided Filter
again (using the same parameters). The base layer is the output of the
filter and contains the coarse image structure, and the detail layer
is the difference with the original image that contains the
high-frequency details. We then apply patch synthesis on the base
layer only and add back the details. This procedure ensures that the
texture is not degraded by the averaging, thereby producing crisp
results.

\section{Painting Estimator}
\label{sec:painting-estimator}

\begin{table}[htp]
\centering
\begin{tabular}{llc}
\toprule
Strength & Art style examples  &  $(\styleweight,\histweight)$  \\ 
\midrule
Weak & Baroque, High Renaissance  & $(1.0, 1.0)$ \\ 
Medium & Abstract Art, Post-Impressionism  & $(5.0, 5.0)$ \\ 
Strong & Cubism, Expressionism & $(10.0, 10.0)$ \\ 
\bottomrule
\end{tabular}
\caption{Weights for selected art styles. Please refer to the
  supplemental document for compact art styles and parameter
  weights. The final weight is a linear interpolation of different art
  styles using our trained painting estimator network. TV weights are computed separately based on the noise level of the painting image (Sec.~\ref{sec:impl-details}) .}
\label{table:weights}

\end{table}

The above algorithm has two important parameters that affect the stylistic
properties of the output --- style and histogram weights ($\styleweight$ and
$\histweight$). We observed that different sets of parameters gave
optimal results for different paintings based on their level of stylization.
For example, Cubism paintings often contain small multifaceted areas with
strong and sharp brush strokes, while High Renaissance and Baroque paintings
are more photorealistic. Rather than tweak parameters for each input, we
 developed a trained predictor of the weights to make our approach to
weight selection more robust.

We  train a \emph{painting estimator} that predicts the optimization parameters for our
algorithm such that parameters that allow deeper style changes are used when
the background painting is more stylized and vice versa. To train this estimator, we split the parameter
values into three categories (``Weak'', ``Medium'' and ``Strong''),
and manually assign each painting style to one of the categories.
Table~\ref{table:weights} presents a subset of painting styles and their
categories and weight values. \fujun{We selected the 18 most common styles based on \texttt{wikiart.org} ranking: Abstract Art, Abstract Expressionism, Art Nouveau (Modern), Baroque, Color Field Painting, Cubism, Early Renaissance, Expressionism, High Renaissance, Impressionism, Mannerism (Late Renaissance), Naive Art (Primitivism), Northern Renaissance, Post-Impressionism, Realism, Surrealism, Symbolism and Ukiyo-e. More details appear in the supplementary material.}

We collected 80,000 paintings from \texttt{wikiart.org} and fine-tuned the VGG-16 network~\cite{simonyan2014very} on classifying 18 different styles. After training, we remove the last classification layer and use weighted linear interpolation on the softmax layer based on style categories to output a floating value for $\styleweight$ and $\histweight$ indicating the level of stylization. These parameter values (shown in Table~\ref{table:weights}) are then used in the optimization.


  \section{Implementation Details}
\label{sec:impl-details}

\begin{figure*}[htp]   
	\centering

	\begin{subfigure}{.166\textwidth}
		\includegraphics[width=0.95\linewidth]{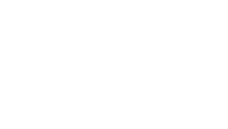}
	\end{subfigure}%
	\begin{subfigure}{.166\textwidth}
		\includegraphics[width=0.95\linewidth]{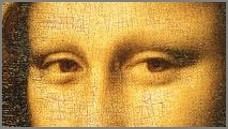}
	\end{subfigure}%
	\begin{subfigure}{.166\textwidth}
		\includegraphics[width=0.95\linewidth]{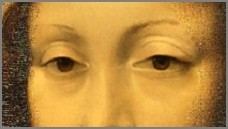}
	\end{subfigure}%
	\begin{subfigure}{.166\textwidth}
		\includegraphics[width=0.95\linewidth]{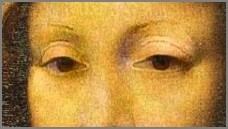}
	\end{subfigure}%
	\begin{subfigure}{.166\textwidth}
		\includegraphics[width=0.95\linewidth]{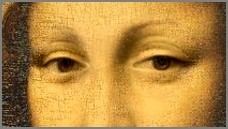}
	\end{subfigure}%
	\begin{subfigure}{.166\textwidth}
		\includegraphics[width=0.95\linewidth]{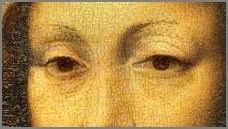}
	\end{subfigure}\vspace{1.2mm}
	
	\begin{subfigure}{.166\textwidth}
		\includegraphics[width=0.95\linewidth]{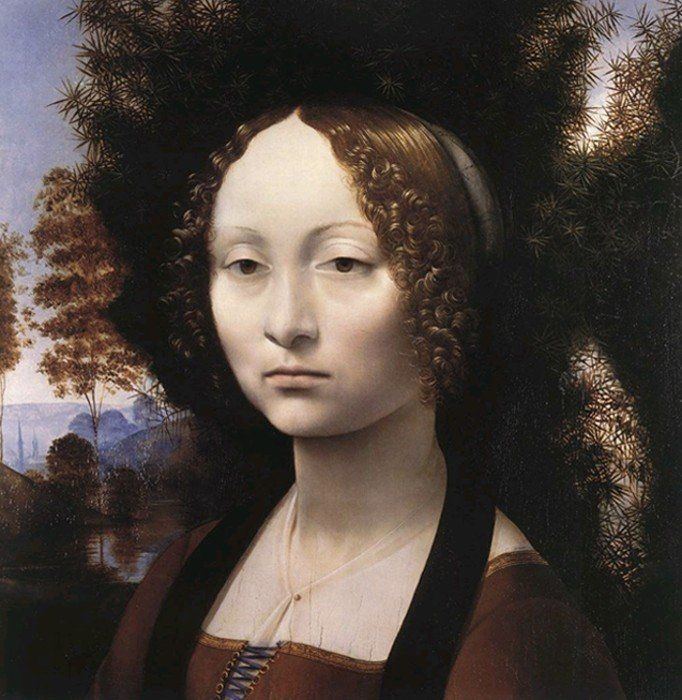}
		\caption{Source}
	\end{subfigure}%
	\begin{subfigure}{.166\textwidth}
		\includegraphics[width=0.95\linewidth]{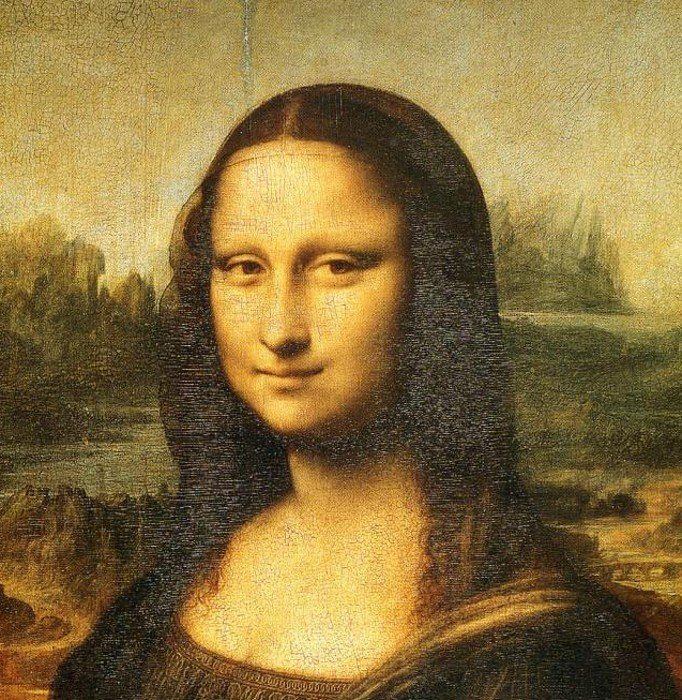}
		\caption{Target painting}
	\end{subfigure}%
	\begin{subfigure}{.166\textwidth}
		\includegraphics[width=0.95\linewidth]{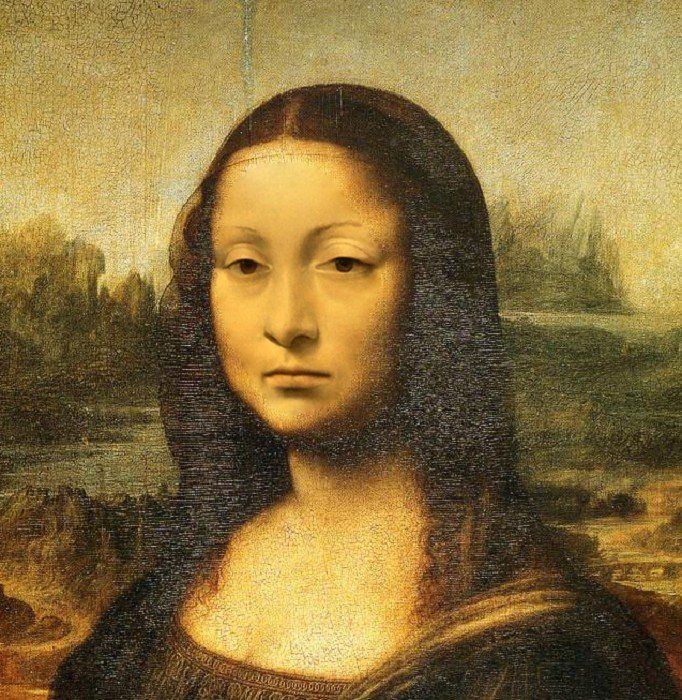}
		\caption{\cite{perez2003poisson}}
	\end{subfigure}%
	\begin{subfigure}{.166\textwidth}
		\includegraphics[width=0.95\linewidth]{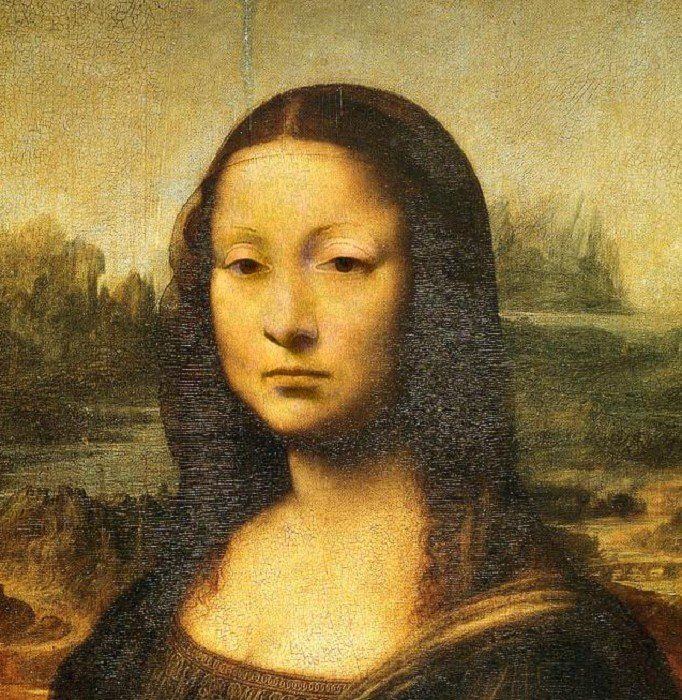}
		\caption{\cite{sunkavalli2010multi}}
	\end{subfigure}%
	\begin{subfigure}{.166\textwidth}
		\includegraphics[width=0.95\linewidth]{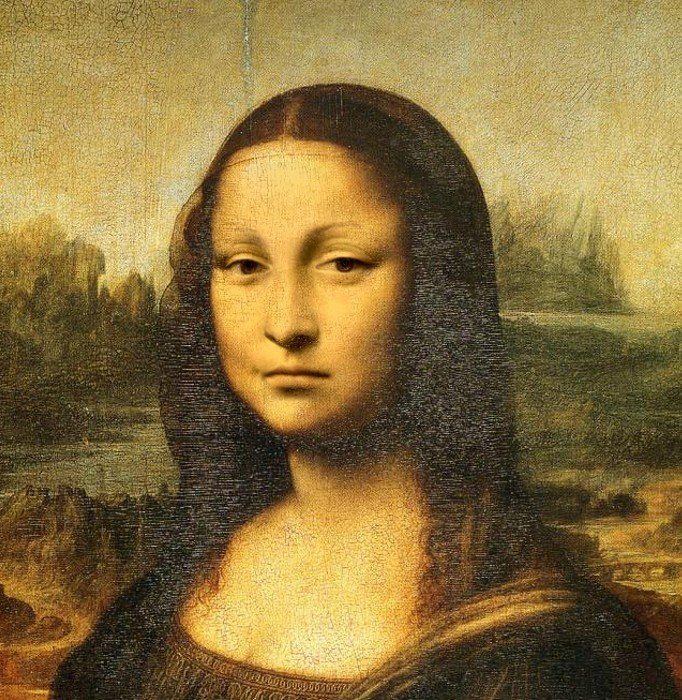}
		\caption{\cite{darabi2012image}}
	\end{subfigure}%
	\begin{subfigure}{.166\textwidth}
		\includegraphics[width=0.95\linewidth]{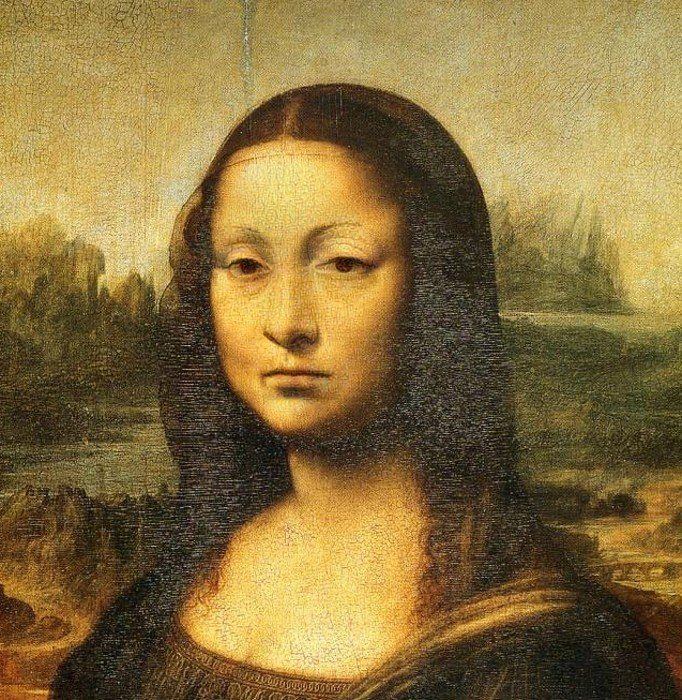}
		\caption{Ours}
	\end{subfigure}
	\vspace{-0.75\baselineskip}
	\caption{
		When pasting the face of Ginevra de' Benci~(a) on Mona Lisa~(b),
		Poisson Blending~\cite{perez2003poisson} does not match the texture~(c), Mulitscale
		Harmonization~\cite{sunkavalli2010multi} adds texture but does not reproduce the paint
		cracks~(d), \fujun{Image Melding~\cite{darabi2012image} adds cracks faithfully but not everywhere, e.g., there are no cracks below the eye on the right~(e). In comparison, our result
		generates cracks more consistently over the image~(f).} 
	}
	\label{fig:monalisa}
	\vspace{-0.75\baselineskip}
\end{figure*}

This section describes the implementation details of our approach. We
employed pre-trained VGG-19~\cite{simonyan2014very} as the feature
extractor. For the first-pass optimization, we chose \vgglayer{conv4\_1}
($\alpha_\ell=1$ for this layer and $\alpha_\ell=0$ for all other
layers) as the content representation, and \vgglayer{conv3\_1},
\vgglayer{conv4\_1} and \vgglayer{conv5\_1} ($\beta_\ell=1/3$ for those
layers and $\beta_\ell=0$ for all other layers) as the style
representation, since higher layers have been transformed by the CNN
into representations with more of the actual content, which is crucial
for the semantic-aware nearest-neighbor search. We used these layer
preferences 
for all the first-pass
results. For the second-pass optimization, we chose \vgglayer{conv4\_1}
as the content representation, \vgglayer{conv1\_1}, \vgglayer{conv2\_1},
\vgglayer{conv3\_1} and \vgglayer{conv4\_1} as the style
representation. We also employed the histogram loss and used
\vgglayer{conv1\_1}, and \vgglayer{conv4\_1} ($\gamma_\ell=1/2$ for these
layers and $\gamma_\ell=0$ for all other layers) as the histogram
representation as suggested by the original authors. We chose
\vgglayer{conv4\_1} as the reference layer $\refneurallayer$ for
the nearest-neighbor search in the second-pass optimization. We name $\tau$ the output
floating number of the painting estimator and  set the 
parameters $\styleweight = \tau$, $\histweight = \tau$, and
$\tvweight = \tau * \texttt{sigmoid}(\texttt{median\_tv}(S))$, where the
$\texttt{sigmoid}(x) = \frac{10}{1 + \exp(10^4 x -
  25)}$
and $\texttt{median\_tv}(S)$ is the median total variation
(Eq.~\ref{eq:tv_loss}) of the painting $S$. We found the parameters
for the sigmoid function empirically. The intuition is that we
impose less smoothness when the original painting is textured.

Our main algorithm is developed in Torch + CUDA. \fujun{We have implemented
  the dense correspondence search and spatial consistency resampling
  in CUDA to manipulate the feature activations from the VGG
  network. The optimization pipeline is based on a popular Torch
  implementation~\footnote{https://github.com/jcjohnson/neural-style}. We
  apply chrominance denoising and patch synthesis  after the optimization pipeline as a separate post-processing step.} All our experiments
are conducted on a PC with an Intel Xeon E5-2686 v4 processor and an
NVIDIA Tesla K80 GPU. We use the L-BFGS solver~\cite{liu1989limited} 
for the reconstruction with 1000 iterations. The runtime on a
$500\times500$ image takes about 5 minutes. We will release our
implementation upon acceptance for non-commercial use and future
research.

\ignorethis{
\paragraph{The Effect of Histogram Loss.} As shown in Figure~\ref{fig:hist_effect}, due to the instabilities and ambiguities of the Gram matrix that there exist multiple output feature maps resulting in the same Gram matrix. It is therefore hard to transfer pointy texture such as the result shown in Figure~\ref{fig:hist_effect}(b). Adding an additional histogram loss resolves this problem.

\paragraph{Addressing CNN Deconvolution Artifacts.} Images deconvolved from
convolutional neural networks can have checkboard artifacts. Also, residual
losses exist even after many iterations during optimization which could lead to
color differences between the foreground and background. These artifacts are
shown in Figure~\ref{fig:cnn_artifact}(a)(e). To fix the checkboard artifacts,
we transform the image to $Lab$ colorspace to filter the $a$ and $b$ channels
using guided filtering~\cite{he2013guided} and reconstruct the output, as shown in
Figure~\ref{fig:cnn_artifact}(d)(h).
 
To remove the residual color artifacts, we use guided filtering to decompose the
image into a detail layer and base layer, and perform \texttt{nn} search in the
painting for the base layer. The output is reconstructed by:
\begin{equation}
O = I - \mathcal{F}(I, G) + \texttt{nn}(\mathcal{F}(I, G), S)
\end{equation}
where $I$ is the direct output from deconvolution optimization, $G$ is the unadjusted cut-and-paste as the guidance, $\mathcal{F}$ is the guided filter, $S$ is the target painting and $O$ is our result after fixing color. The nearest neighbor, \texttt{nn}, search can be accelerated using PatchMatch~\cite{barnes2009patchmatch}.

}

\ignorethis{
\kb{Note that deconvolution artifact issues have been mentioned
in other related work~\cite{odena2016deconvolution}, where the reason is the uneven
overlapping kernels in the deconvolution. One straightforward solution proposed has been to
redesign the neural network architecture carefully to balance the overlapping
kernels. Unfortunately, most existing style transfer frameworks rely on the
pre-trained VGG network which means the structure is fixed. One possible
approach is to modify the VGG network structure and retrain it for style
transfer. Alternatively, our developed strategy above addresses these artifacts
satisfactorily.}
}


  \section{Results}

We now evaluate our harmonization algorithm in comparison with related work and
through user studies.

\paragraph{Main Results.} In
Figures~\ref{tab:main_res1} and ~\ref{tab:main_res2}, we compare our method with four state-of-the-art
methods: Neural Style~\cite{gatys2015neural},
CNNMRF~\cite{li2016combining}, Multi-Scale Image
Harmonization~\cite{sunkavalli2010multi}, and Deep Image
Analogy~\cite{liao2017visual} across paintings with various styles and
genres.  Neural Style tends to produce ``style summaries'' that rarely
work well; for example, background sky texture appears in the foreground eiffel tower
(Fig.~\ref{tab:main_res1}(iv)). This is due
to the lack of semantic matching since the Gram matrix is computed
over the entire painting. CNNMRF often generates weak style transfers
that do not look as good when juxtaposed with the original
painting. Multi-Scale Image Harmonization performs noise matching to
fit high-frequency inter-scale texture but does not capture
spatially-varying brush strokes common in paintings with heavy
styles. Deep Image Analogy is more robust compared to the other three
methods but its results are sometimes blurred due to patch synthesis
and voting, e.g.,
Figures~\ref{tab:main_res1}(i-iii),~\ref{tab:main_res2}(vii). Its
coarse-to-fine pipeline also sometimes misses parts
(Fig.~\ref{tab:main_res1}(iv)).


\paragraph{User Studies.} We conduct user studies to quantitatively
characterize the quality of our results.  The first user study,
``Edited or Not'', aims to understand whether the harmonization
quality is good enough to fool an observer. The second user study,
``Comparison'', compares the quality of our results with that of
related algorithms.

\begin{figure}[htp]
\centering
\includegraphics[width=1\linewidth]{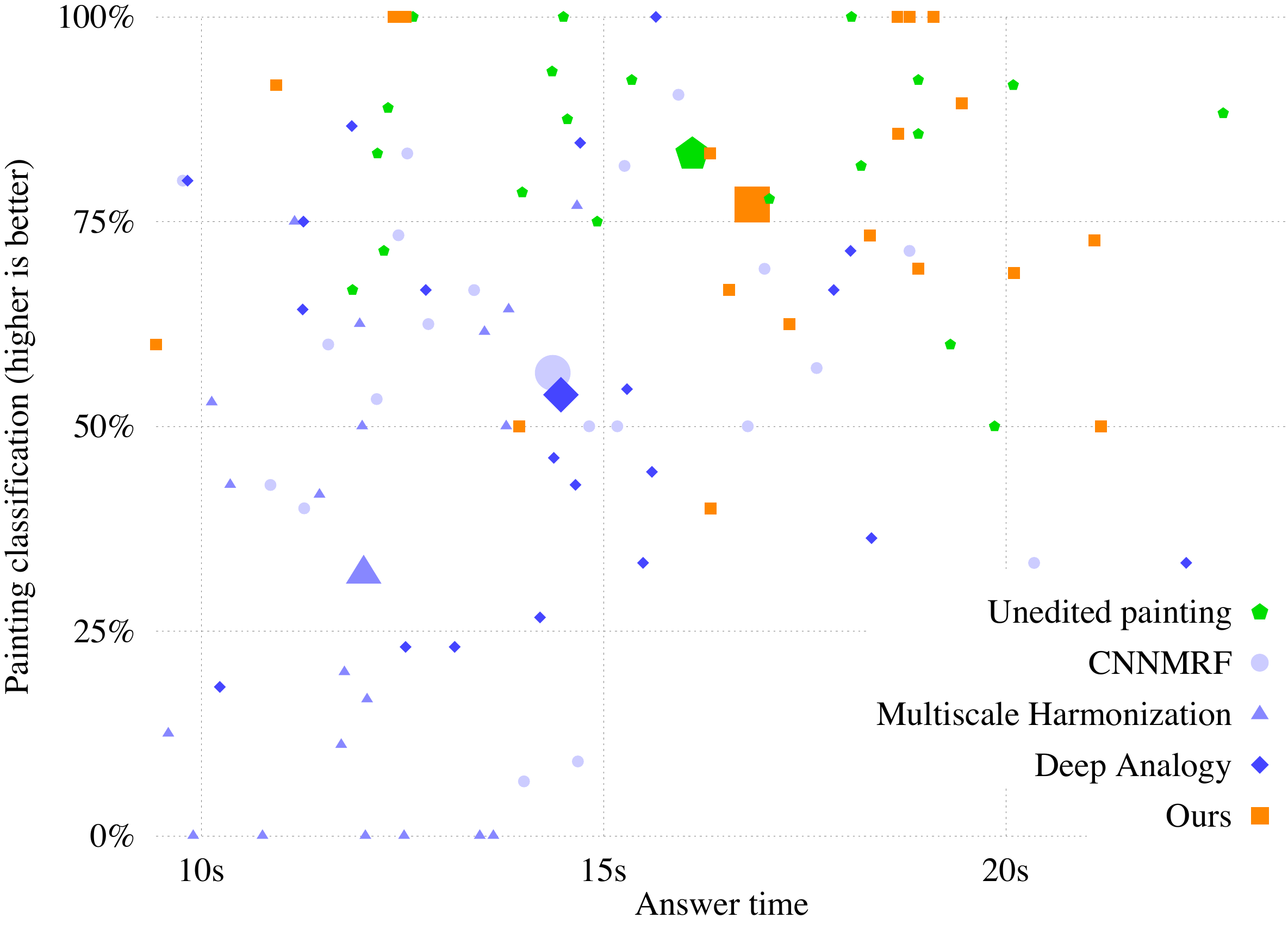}
\caption{Results of the``Edited or Not'' user study. A higher
  painting classification rate means better harmonization performance since users
  were unable to identify the edit. See text for more details. The
large symbols represent the average of each category.}
\label{fig:survey1}
\end{figure}

\noindent{\bf Study 1: Edited or Not.} 
We showed the users 20 painterly composites, each edited by one of
four algorithms: CNNMRF~\cite{li2016combining}, Multi-Scale Image
Harmonization~\cite{sunkavalli2010multi}, Deep Image
Analogy~\cite{liao2017visual}, and ours. We asked the users whether
the painting had been edited. If they thought it was, we asked them to
click on the part of the image they believed was edited (this records
the coordinates of the edited object) so that we could verify the
correctness of the answer. This verification is motivated by our pilot study
where we found that people would sometimes claim an image is edited by
erroneously identifying an element as edited although it was part of
the original painting.  Such misguided classification is actually a
positive result that shows the harmonization is of high quality.
We also recorded the time it took users to answer each question.

One potential problem in the study that we had to consider is that
people might spot the edited object in a painting due to reasons other
than the harmonization quality. For example, an edit in a famous
painting will be instantly recognizable, or if the composition of the
edited painting is semantically wrong, e.g., a man's face in a woman's head, or a spaceship in a painting from the 19th century.  To avoid these problems,
we selected typically unfamilar paintings and made the composition
sensible; for example adding a park bench in a meadow or a clock on a
wall (see supplementary material for more examples). We further asked
the users for each example if they were familiar with the painting.
If they were, we eliminated their judgement as being tainted by prior
knowledge.

Figure~\ref{fig:survey1} shows the results of Study 1 using two
metrics: the average \emph{painting classification rate} and average
\emph{answer time}.  Let $N(x)$ denote the total number of users
with answer $x$.  For an original painting, the painting
classification rate is computed as $N_n / (N_n + N_e)$, where $N_n$ is
the number of answers with $\mathtt{Not\ Edited }$ and $N_e$ is the
number of answers with $\mathtt{Edited}$. For edited paintings using
those four algorithms, the painting classification rate $P$ is
computed as $(N_n +\hat{ N_{e} }) / (N_n + N_e)$, where $\hat{N_{e}}$
is the number of answers with $\mathtt{Edited}$ but with the wrong XY
coordinates of the mouse click.  This captures all the cases where the
viewer was ``fooled'' by the harmonization result.  A higher 
rate means better harmonization quality since users were unable to
identify the modification. Figure~\ref{fig:survey1} shows that our
algorithm achieves a painting classification rate significantly higher than
that of the other algorithms and close to that of unedited paintings.
%

The answer time has a less straightforward interpretation since it may
also reflect how meticulous users are. Nonetheless,
Figure~\ref{fig:survey1} shows that the answer time for our algorithm is
close to that of unedited paintings and significantly different from
that of the other algorithms, which also suggests that our results
share more similarities with actual paintings than the outputs of the
other methods.


\begin{figure}[htp]
\centering
\includegraphics[width=\linewidth]{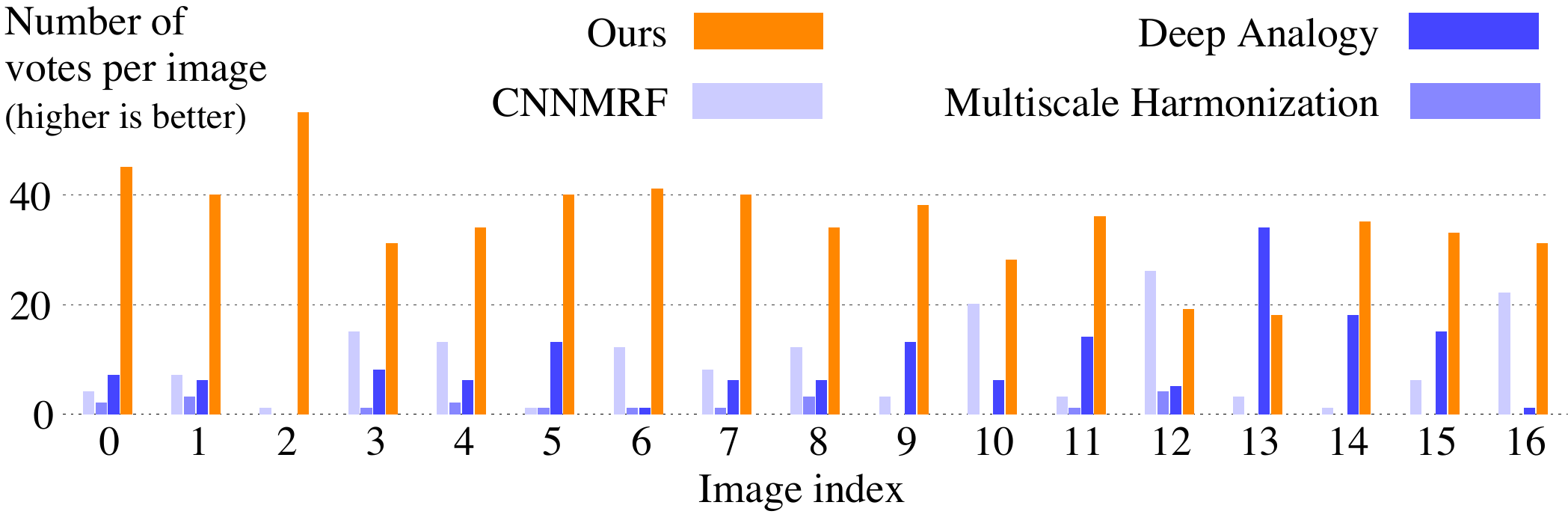}
\caption{Results of the ``Comparison'' user study. Our algorithm is
  often the most preferred among the four algorithms.} 
\label{fig:survey2}
\end{figure}

\noindent{\bf Study 2: Comparison.} We showed the users 17 paintings,
each painting had been edited with the same four
algorithms as the first  study. We asked the users to select the
result that best captures the consistency of the colors and of the texture in
the painting. The quantitative results are shown in Figure~\ref{fig:survey2}.
For most paintings our algorithm is most preferred.  We provide more detailed
results in the supplementary document.

\paragraph{Image Harmonization Comparisons.} We compare our results
with Poisson blending and two state-of-the-art harmonization
solutions, \cite{sunkavalli2010multi} and \cite{darabi2012image} in
Figure~\ref{fig:monalisa}. Poisson blending achieves good overall
color matching but does not capture the texture of the original painting.
Multiscale Harmonization~\cite{sunkavalli2010multi}
transforms noise to transfer texture properties, in addition to color
and intensity. However, it is designed to fit small-scale, noise-like texture and is not well suited for more structured patterns
such as painting brush strokes and cracks. Image Melding~\cite{darabi2012image}
improves texture quality by using patch synthesis combined with Poisson blending, but the texture disappears at
some places.  
In comparison, our method better captures both the spatial and
inter-scale texture and structure properties
(Fig.~\ref{fig:monalisa}f).

\begin{figure}[htp]
\centering
\includegraphics[width=0.32\linewidth]{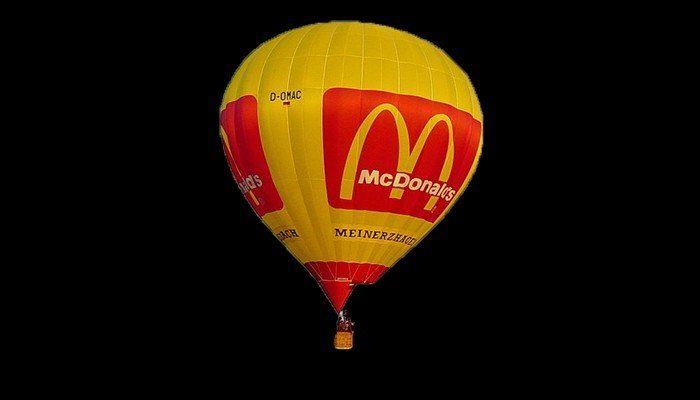}\vspace{0.2mm}
\includegraphics[width=0.32\linewidth]{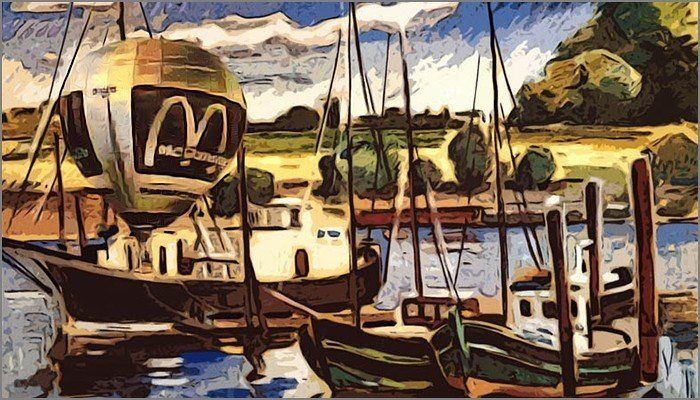}\vspace{0.2mm}
\includegraphics[width=0.32\linewidth]{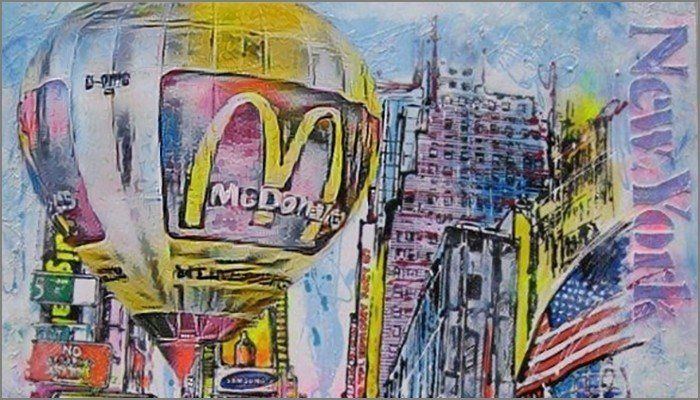}\vspace{0.2mm}

\includegraphics[width=0.32\linewidth]{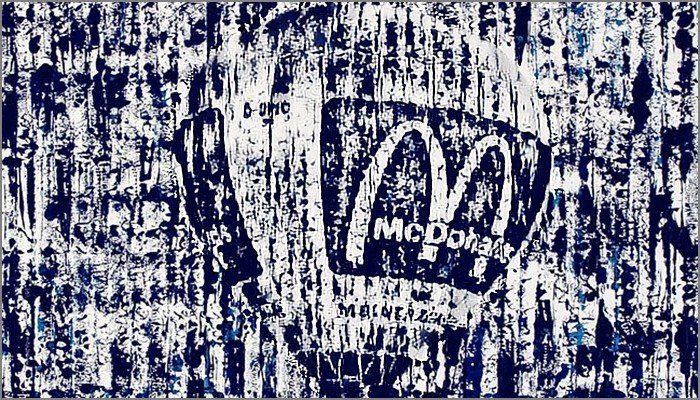}\vspace{0.2mm}
\includegraphics[width=0.32\linewidth]{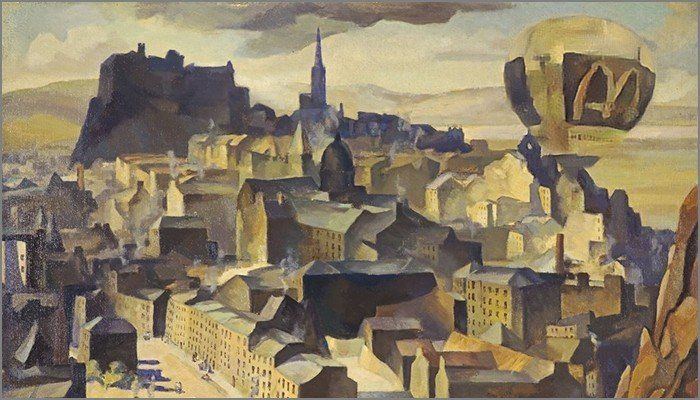}\vspace{0.2mm}
\includegraphics[width=0.32\linewidth]{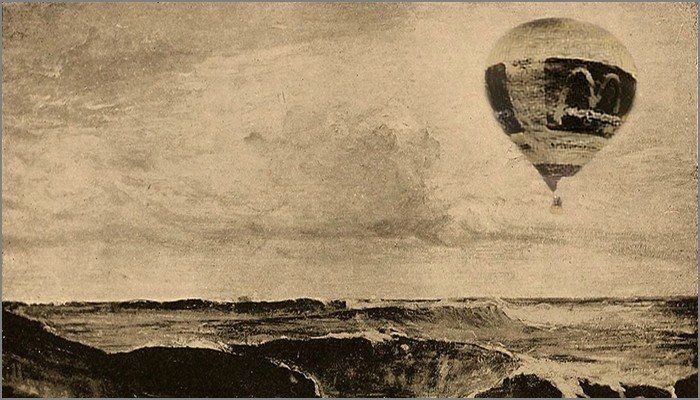}\vspace{0.2mm}

\includegraphics[width=0.32\linewidth]{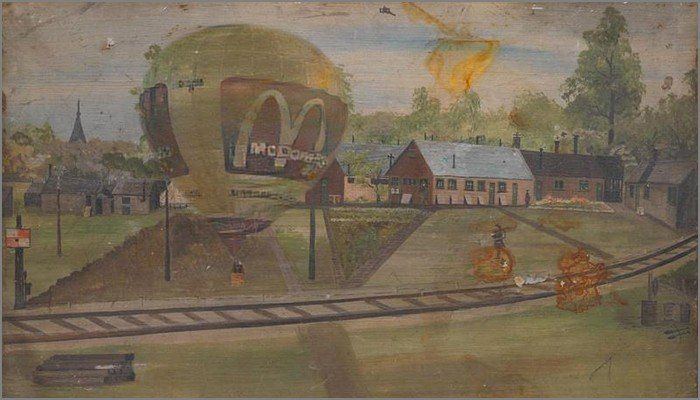}\vspace{0.2mm}
\includegraphics[width=0.32\linewidth]{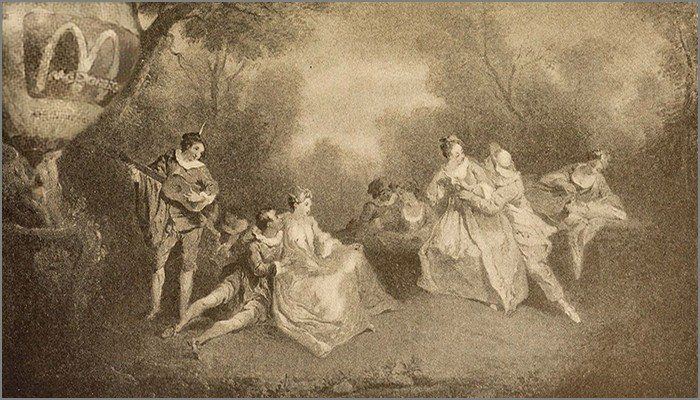}\vspace{0.2mm}
\includegraphics[width=0.32\linewidth]{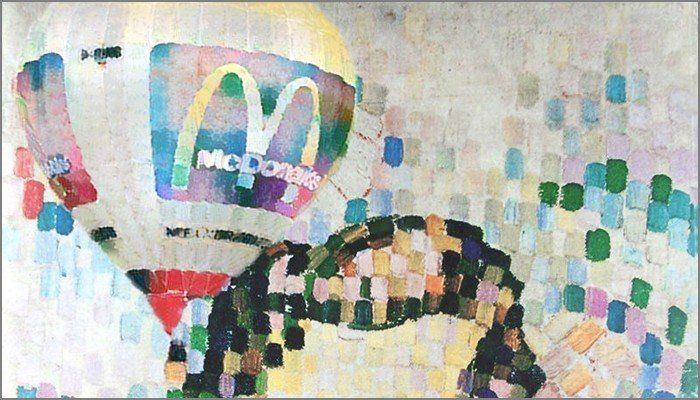}\vspace{0.2mm}

\caption{Canonical object harmonization results for hot air balloon (upper-left).}

\label{fig:star_trek}
\end{figure}

\paragraph{Harmonization of a canonical object across styles.} In the above
examples, we picked different objects for different paintings to create plausible
combinations.
In this experiment, we use the same canonical object 
across a variety of styles to demonstrate the stylization independent of
the inserted object.  We introduce a hot air balloon 
into paintings with a wide range of styles.  We randomly selected paintings
from the \texttt{wikiart.org} dataset to paste the balloon into
(Fig.~\ref{fig:star_trek}). \fujun{Note that some of the results are stylized
so strongly that they are difficult to distinguish from the background
painting. This is similiar to Camouflage Images~\cite{chu2010camouflage}, and is
a limitation of our approach and an interesting future research direction.}





  \vspace{-0.4cm}
\section{Conclusions}

We have described an algorithm to copy an object in a photograph and paste
it into a painting seamlessly, i.e., the composite still looks like a
genuine painting. We have introduced a two-pass algorithm that first
transfers the overall style of the painting to the input and then
refines the result to accurately match the painting's color and
texture. This latter pass relies on mapping neural response statistics
that ensures consistency across the network layers and in image space. To
cope with different painting styles, we have trained a separate
network to adjust the transfer parameters as a function of the
style of the background painting. Our experiments show that our
approach succeeds on a diversity of input and style images, many of
which are challenging for other methods. We have also conducted two
user studies that show that users often identify our results as
unedited paintings and prefer them to the outputs of other
techniques.

We believe that our work opens new possibilities for creatively editing and
combining images and hope that it will inspire artists. From a technical
perspective, we have demonstrated that global painterly style transfer methods
are not well suited for local transfer, and we 
designed an effective
local approach. This suggests fundamental differences between the local and
global statistics of paintings, and
further exploring this difference is an exciting avenue for future work.
Other avenues of future work include fast feed-forward network approximations of our optimization framework, as well an extension to painterly video compositing.

\ignorethis{
We introduce a deep-learning approach that harmonizes object-painting compositings for a wide variety of art styles. We propose a painting estimator which effectively estimates the level of stylization of a given painting image by fine-tuning pre-trained VGG-16 on a large artwork dataset collected from \texttt{wikiart.org}. We also propose a two-step optimization which produces semantic-aware harmonization results and outperforms the state-of-the-arts. 
}

\begin{figure*}[htp]
\begin{adjustwidth}{-0.45cm}{}

\begin{tabular}{ c | c }
 
\multirow{2}{*}{(i)}
& 
	\includegraphics[width=.162\textwidth]{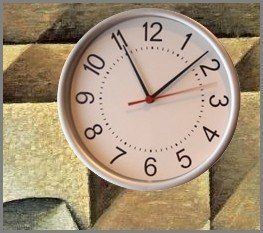}
	\includegraphics[width=.162\textwidth]{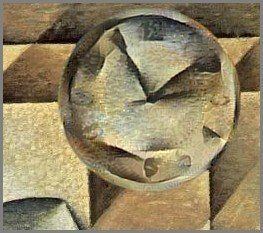}
	\includegraphics[width=.162\textwidth]{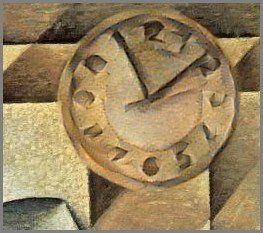}
	\includegraphics[width=.162\textwidth]{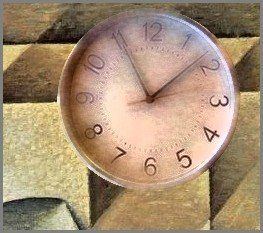}
	\includegraphics[width=.162\textwidth]{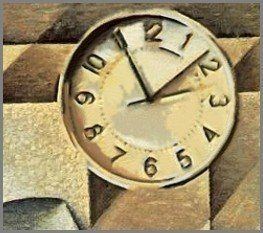}
	\includegraphics[width=.162\textwidth]{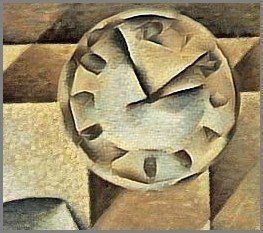}
\\ 
&
	\includegraphics[width=.162\textwidth]{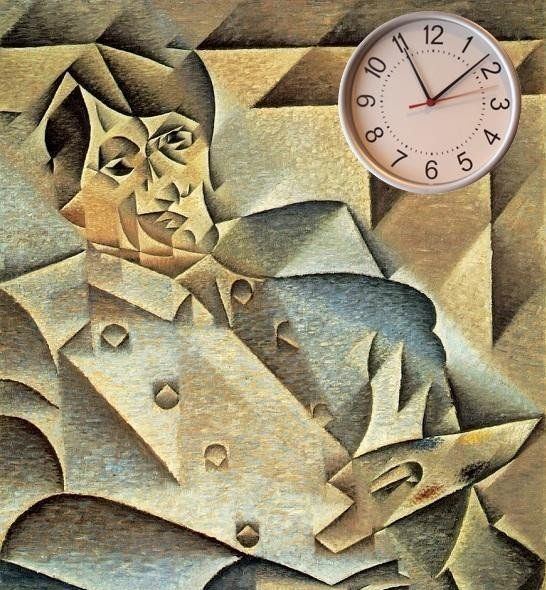}
	\includegraphics[width=.162\textwidth]{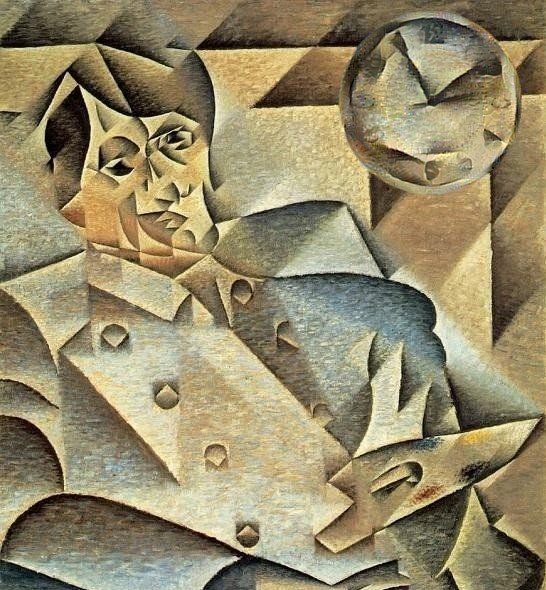}
	\includegraphics[width=.162\textwidth]{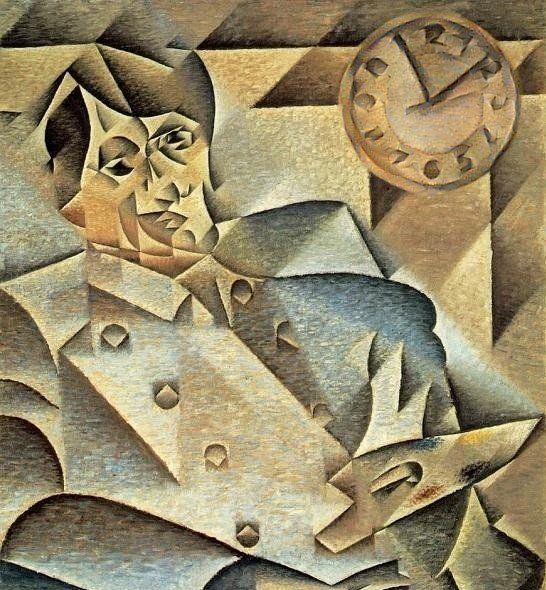}
	\includegraphics[width=.162\textwidth]{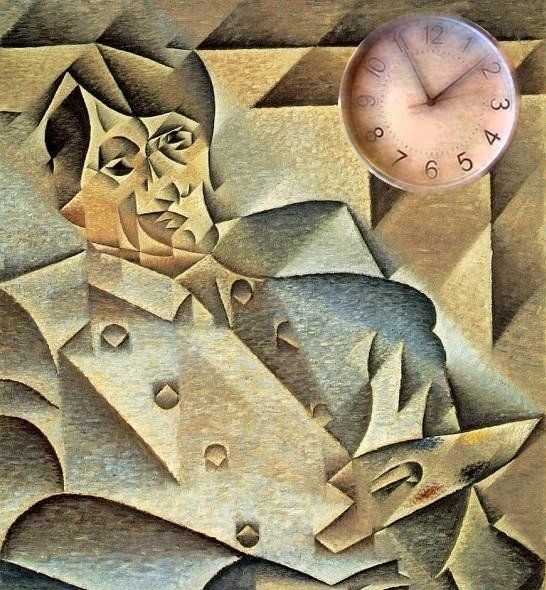}
	\includegraphics[width=.162\textwidth]{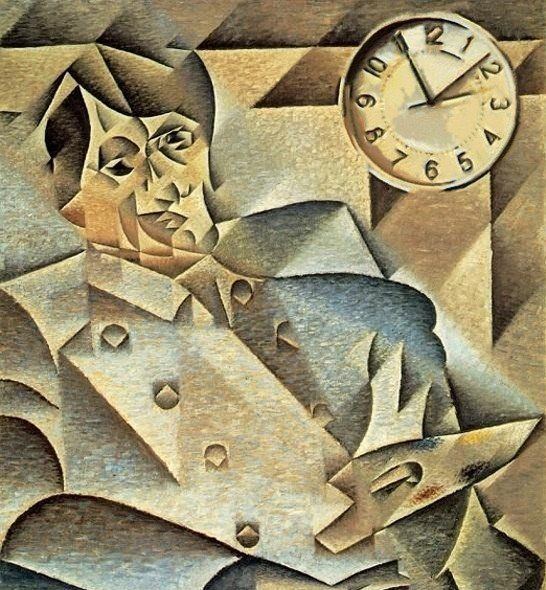}
	\includegraphics[width=.162\textwidth]{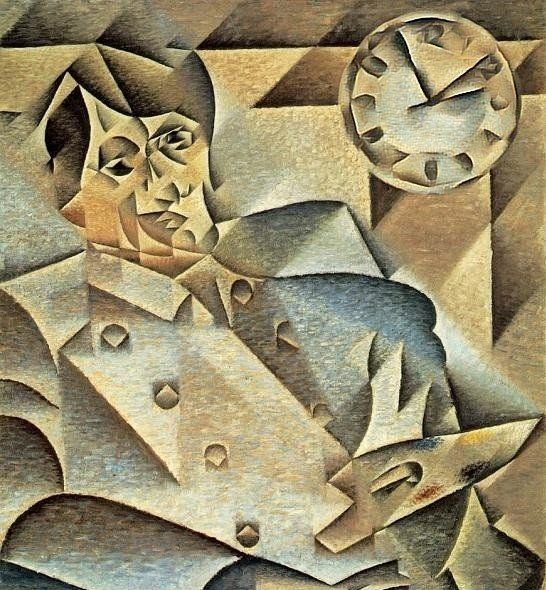}

\\  \hline \\ [-2ex]

\multirow{2}{*}{(ii)}
& 
	\includegraphics[width=.162\textwidth]{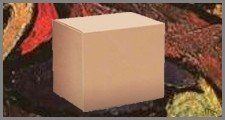}
	\includegraphics[width=.162\textwidth]{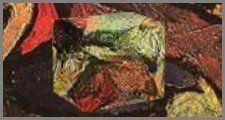}
	\includegraphics[width=.162\textwidth]{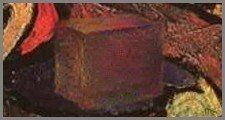}
	\includegraphics[width=.162\textwidth]{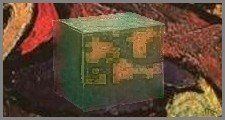}
	\includegraphics[width=.162\textwidth]{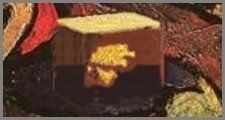}
	\includegraphics[width=.162\textwidth]{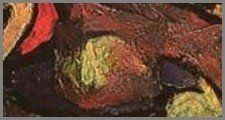}
\\  
&
	\includegraphics[width=.162\textwidth]{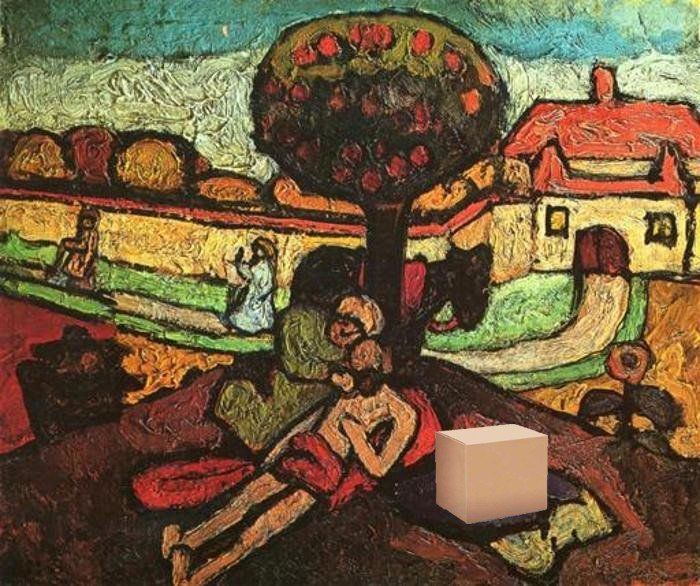}
	\includegraphics[width=.162\textwidth]{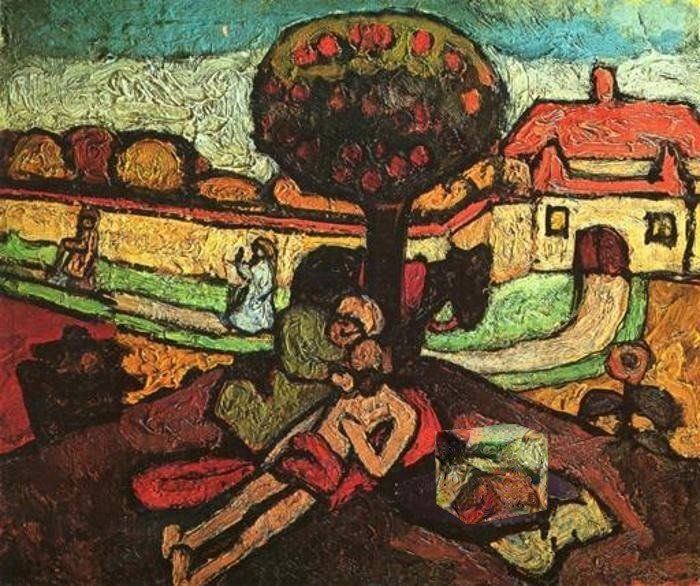}
	\includegraphics[width=.162\textwidth]{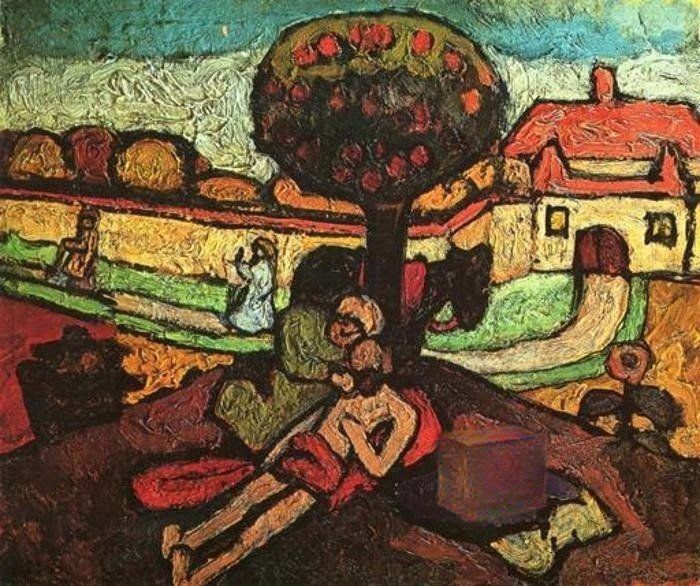}
	\includegraphics[width=.162\textwidth]{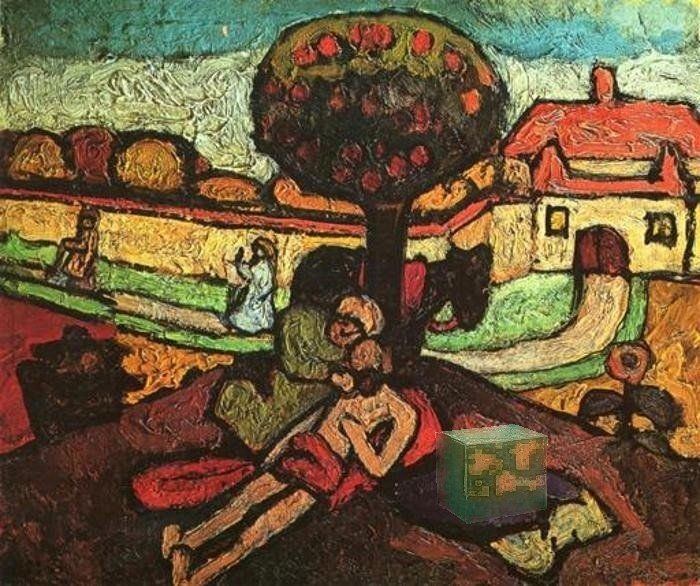}
	\includegraphics[width=.162\textwidth]{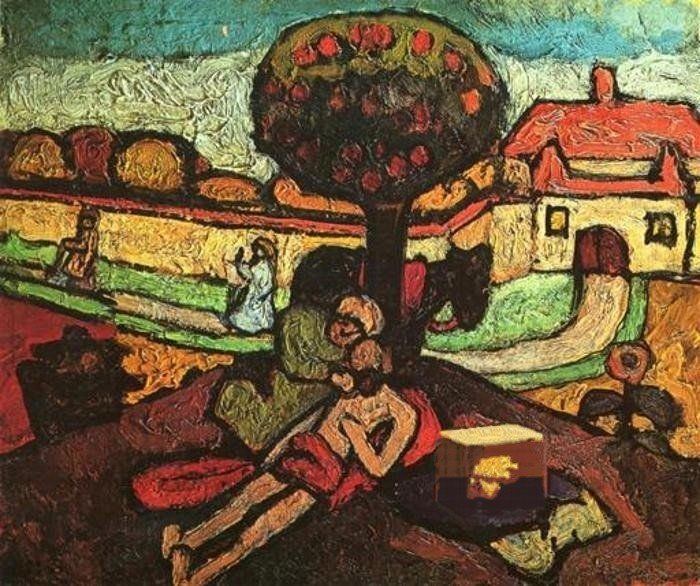}
	\includegraphics[width=.162\textwidth]{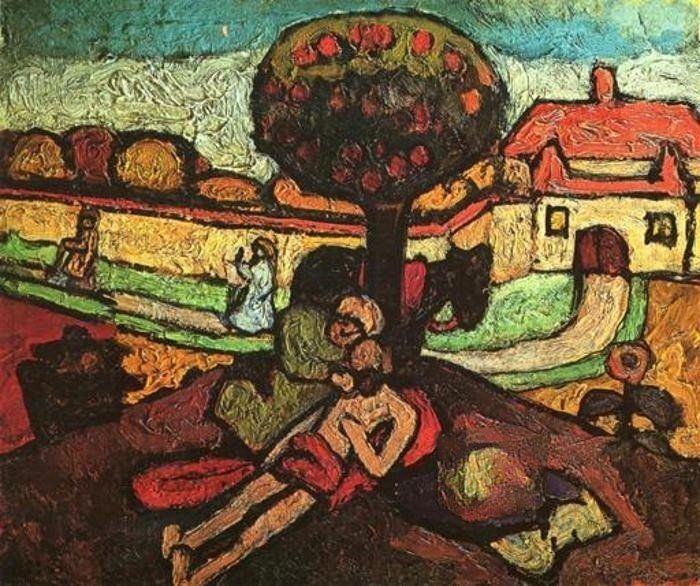}

\\  \hline \\ [-2ex]

\multirow{2}{*}{(iii)}
&
	\includegraphics[width=.162\textwidth]{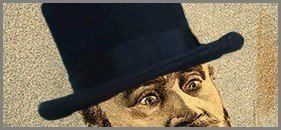}
	\includegraphics[width=.162\textwidth]{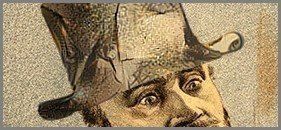}
	\includegraphics[width=.162\textwidth]{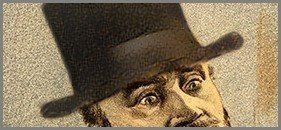}
	\includegraphics[width=.162\textwidth]{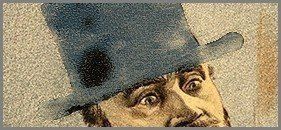}
	\includegraphics[width=.162\textwidth]{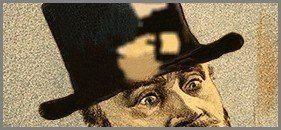}
	\includegraphics[width=.162\textwidth]{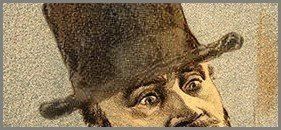}
\\ 
&
	\includegraphics[width=.162\textwidth]{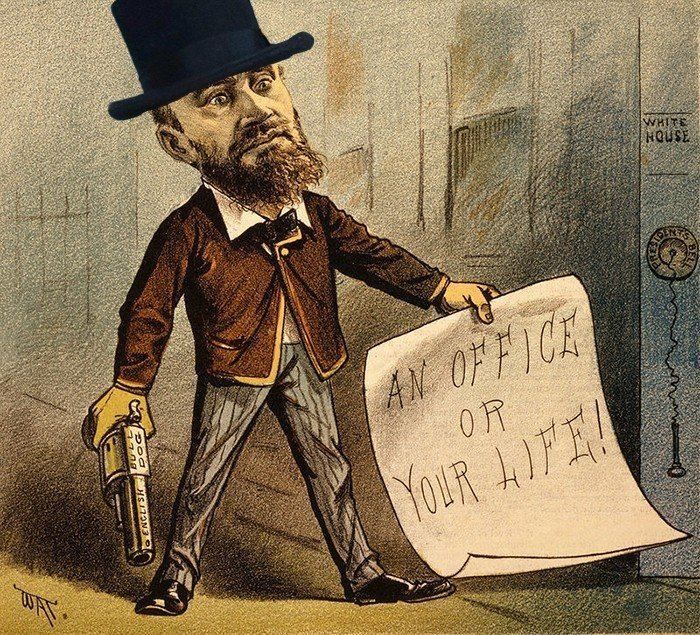}
	\includegraphics[width=.162\textwidth]{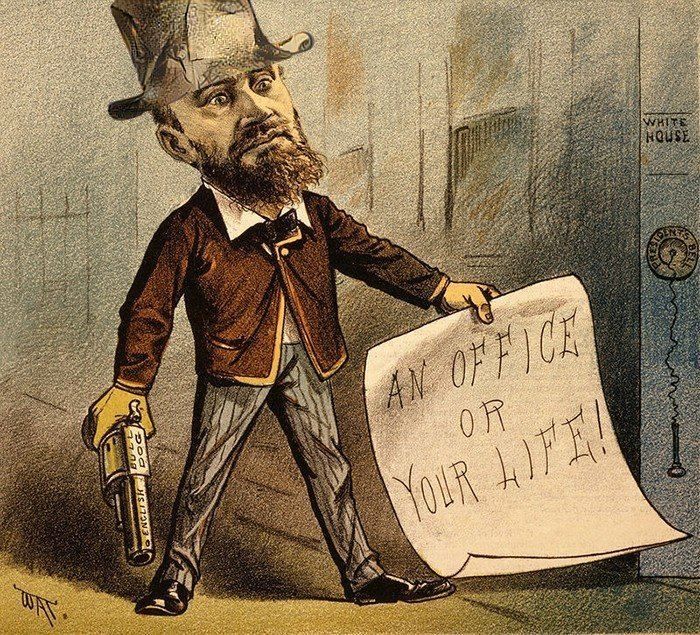}
	\includegraphics[width=.162\textwidth]{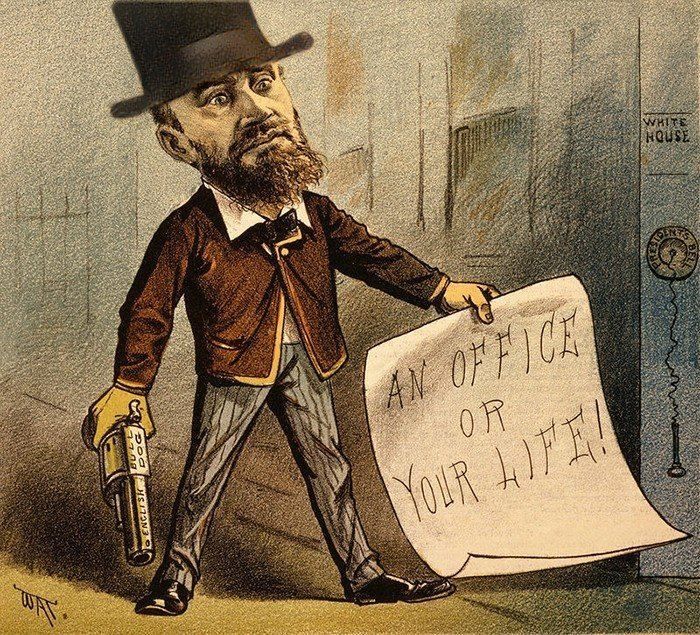}
	\includegraphics[width=.162\textwidth]{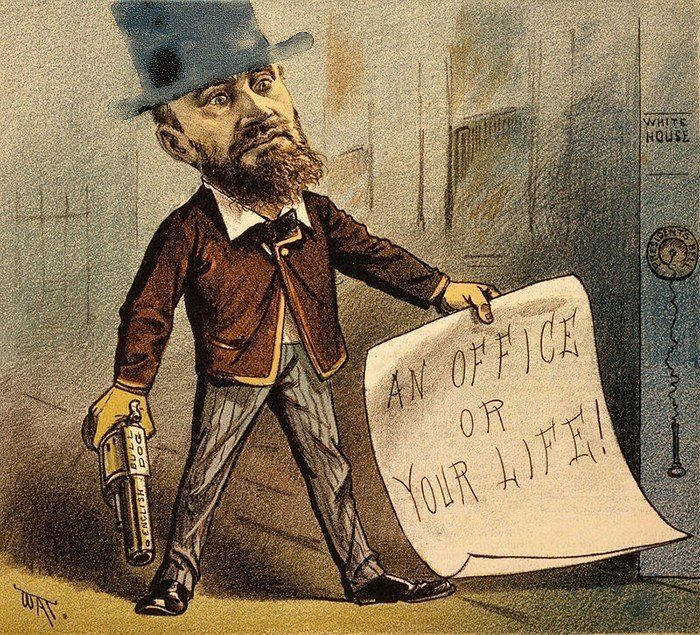}
	\includegraphics[width=.162\textwidth]{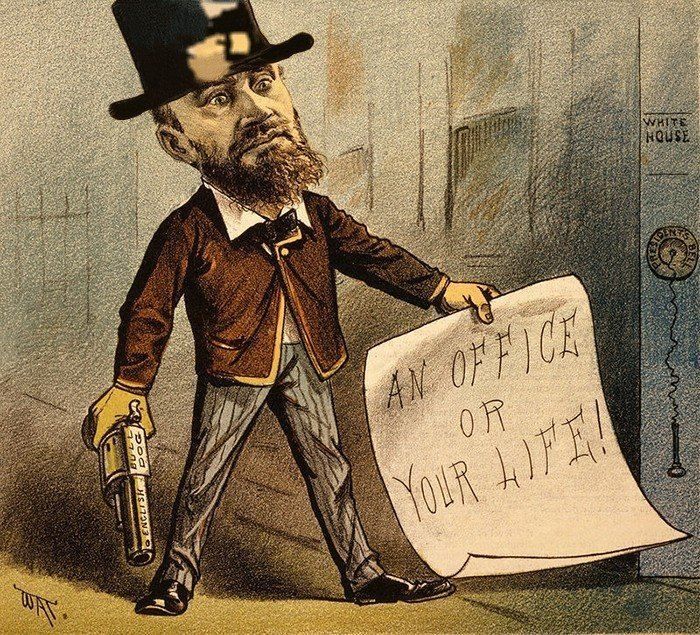}
	\includegraphics[width=.162\textwidth]{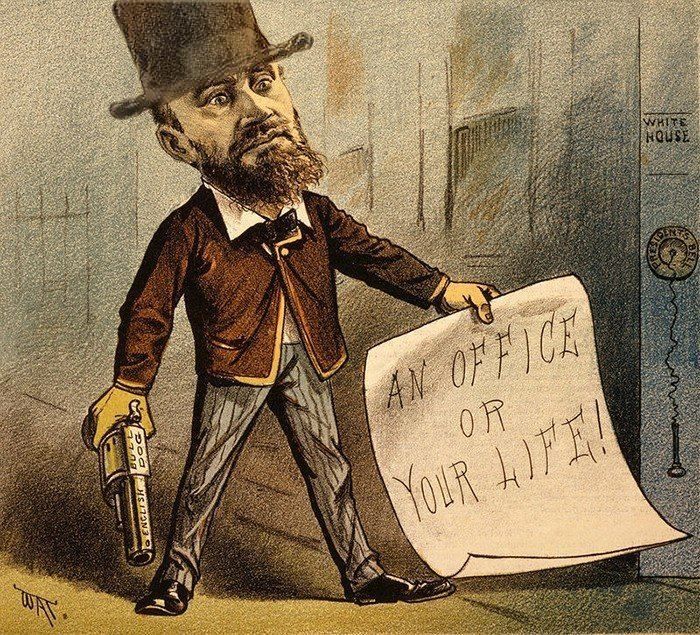}

\\  \hline \\ [-2ex]

\multirow{2}{*}{(iv)}
& 
	\includegraphics[width=.162\textwidth]{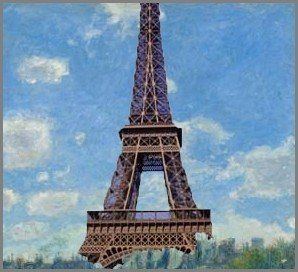}
	\includegraphics[width=.162\textwidth]{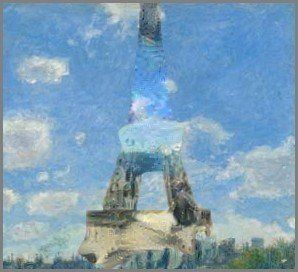}
	\includegraphics[width=.162\textwidth]{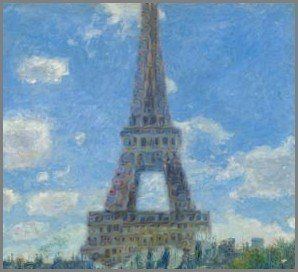}
	\includegraphics[width=.162\textwidth]{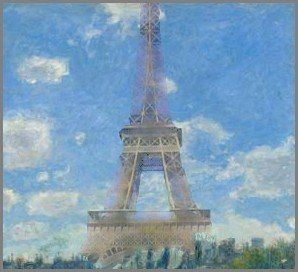}
	\includegraphics[width=.162\textwidth]{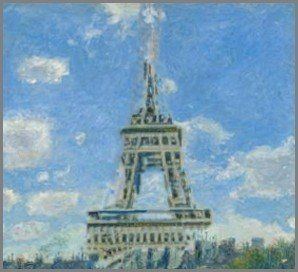}
	\includegraphics[width=.162\textwidth]{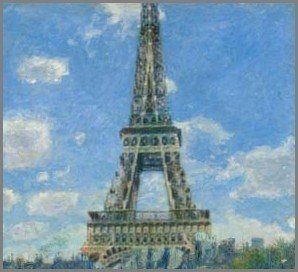}
\\  
&	
	\begin{subfigure}{.162\textwidth}
		\includegraphics[width=1.\linewidth]{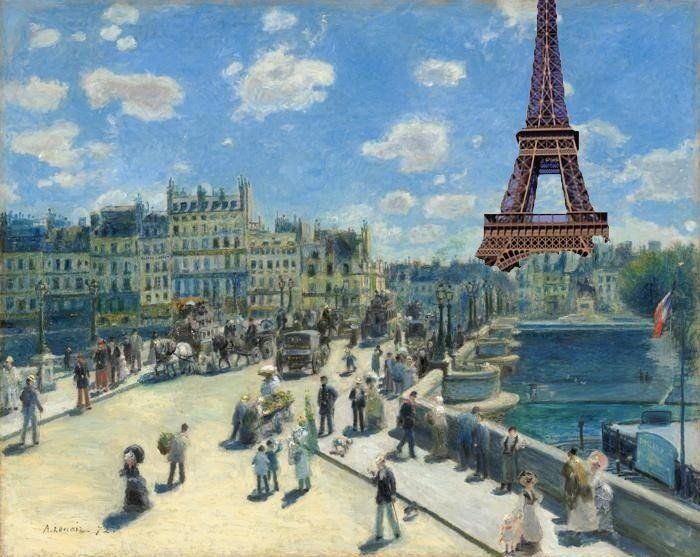}
		\caption{Cut-and-paste}
	\end{subfigure} 
	\begin{subfigure}{.162\textwidth}
		\includegraphics[width=1.\linewidth]{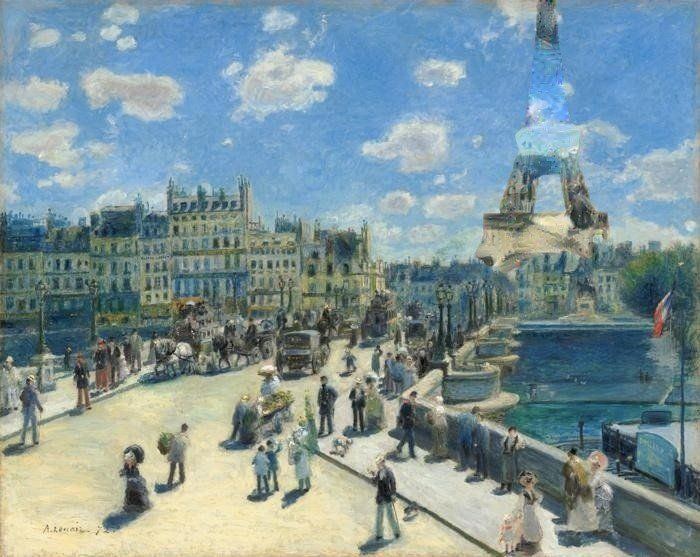}
		\caption{\cite{gatys2015neural}}
	\end{subfigure} 
	\begin{subfigure}{.162\textwidth}
		\includegraphics[width=1.\linewidth]{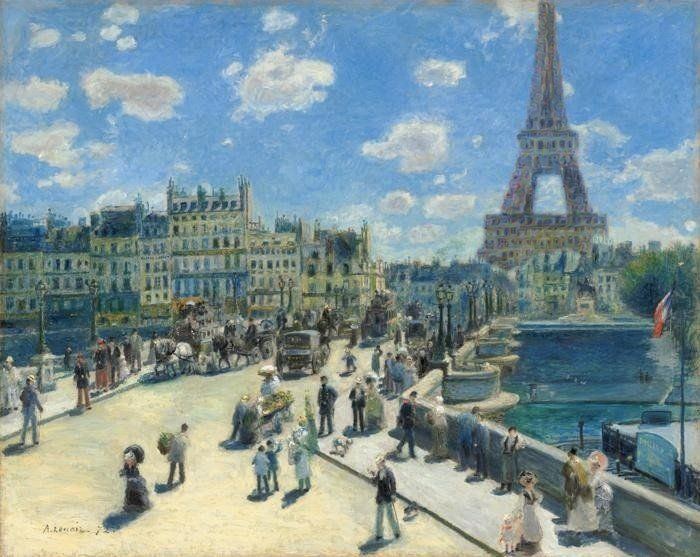}
		\caption{\cite{li2016combining}}
	\end{subfigure}
	\begin{subfigure}{.162\textwidth}
		\includegraphics[width=1.\linewidth]{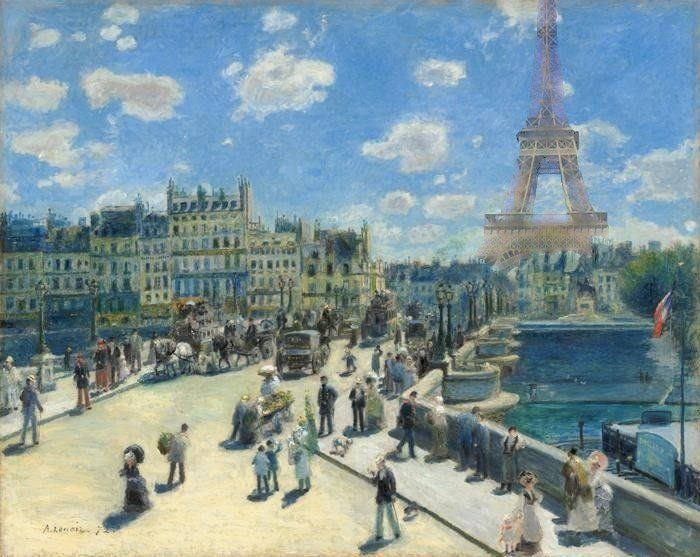}
		\caption{\cite{sunkavalli2010multi}}
	\end{subfigure}
	\begin{subfigure}{.162\textwidth}
		\includegraphics[width=1.\linewidth]{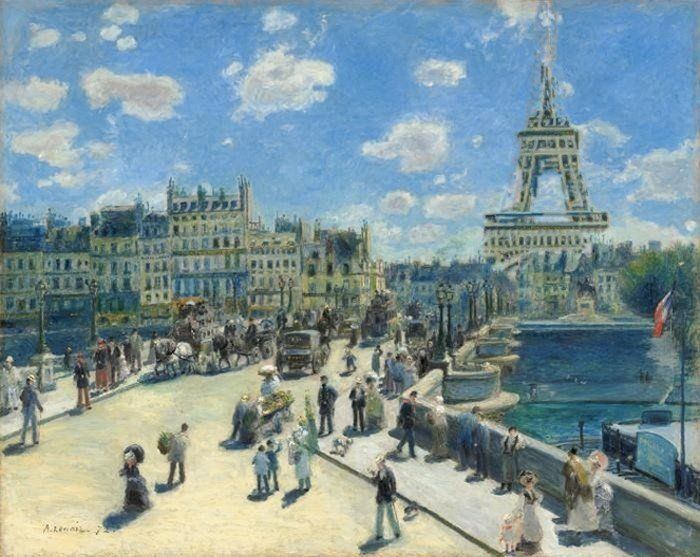}
		\caption{\cite{liao2017visual}}
	\end{subfigure}
	\begin{subfigure}{.162\textwidth}
		\includegraphics[width=1.\linewidth]{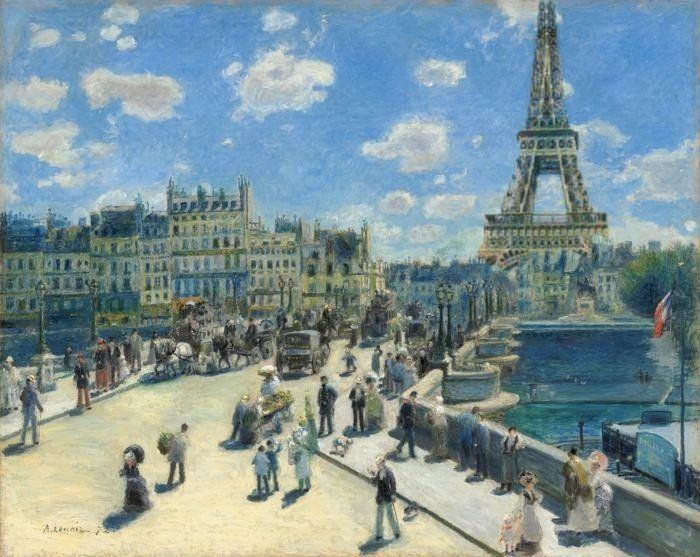}
		\caption{Ours}
	\end{subfigure}

\\   

\end{tabular}

\end{adjustwidth}

\caption{Example results with insets on proposed composite for unadjusted cut-and-paste, four state-of-the-art methods and our results. We show that our method captures both spatial and inter-scale color and texture and produces harmonized results on paintings with various styles. Zoom in for details. }

\label{tab:main_res1}

\end{figure*}

\begin{figure*}[htp]
\begin{adjustwidth}{-0.45cm}{}

\begin{tabular}{ c | c }

\multirow{2}{*}{(v)}
& 
	\includegraphics[width=.162\textwidth]{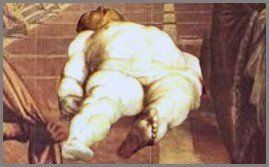}
	\includegraphics[width=.162\textwidth]{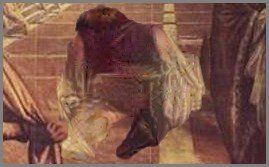}
	\includegraphics[width=.162\textwidth]{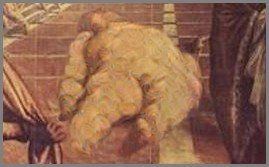}
	\includegraphics[width=.162\textwidth]{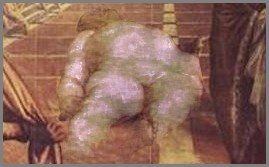}
	\includegraphics[width=.162\textwidth]{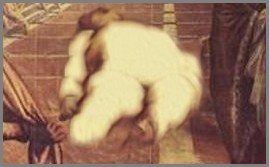}
	\includegraphics[width=.162\textwidth]{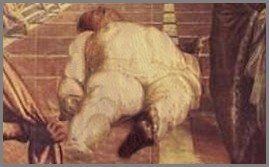}
\\ 
&
	\includegraphics[width=.162\textwidth]{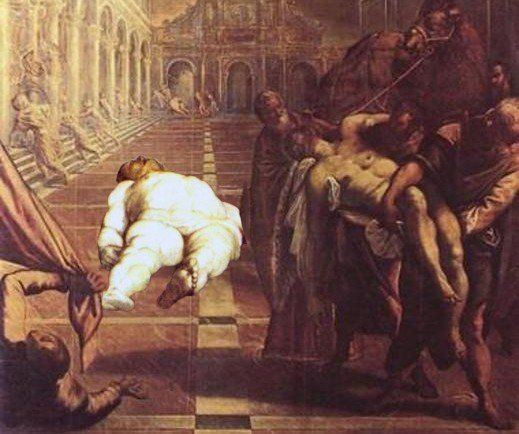}
	\includegraphics[width=.162\textwidth]{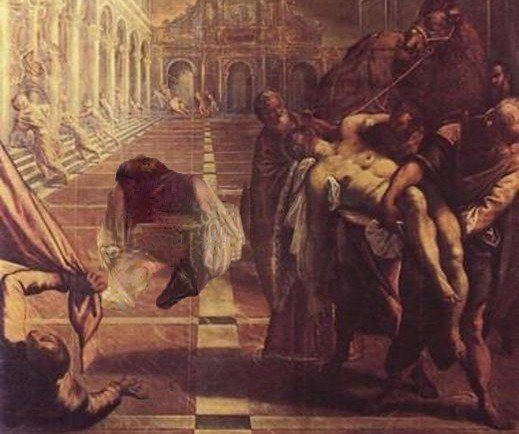}
	\includegraphics[width=.162\textwidth]{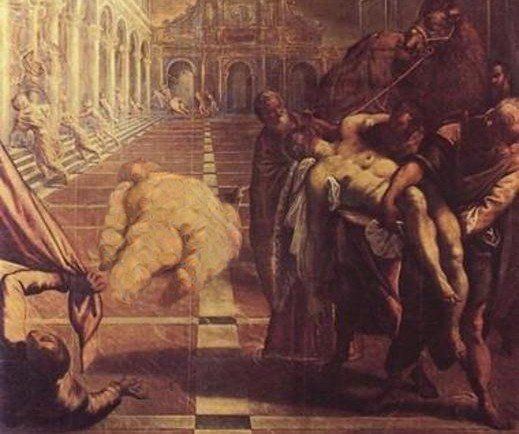}
	\includegraphics[width=.162\textwidth]{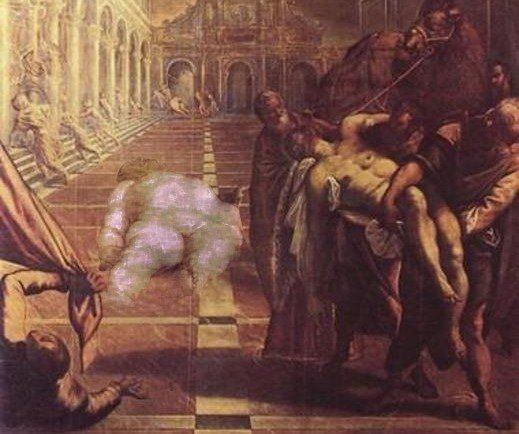}
	\includegraphics[width=.162\textwidth]{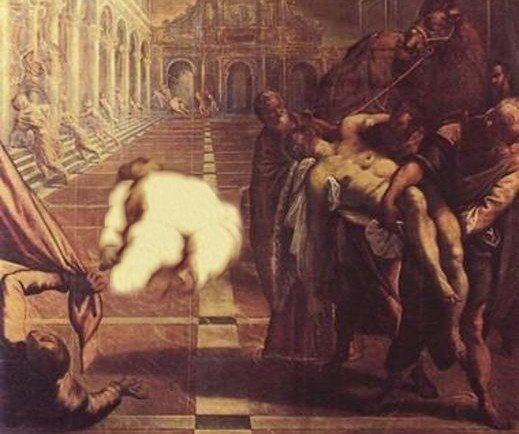}
	\includegraphics[width=.162\textwidth]{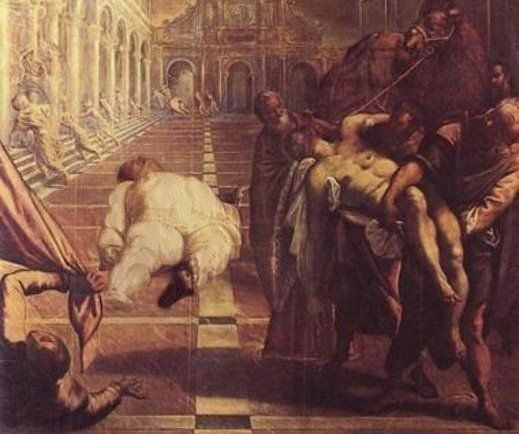}

\\  \hline \\ [-2ex]

\multirow{2}{*}{(vi)}
& 
	\includegraphics[width=.162\textwidth]{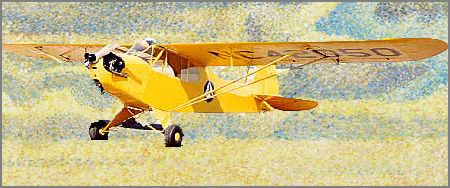}
	\includegraphics[width=.162\textwidth]{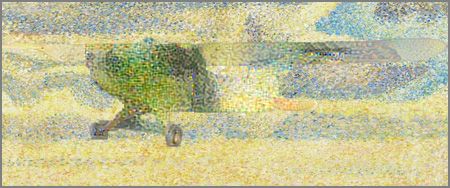}
	\includegraphics[width=.162\textwidth]{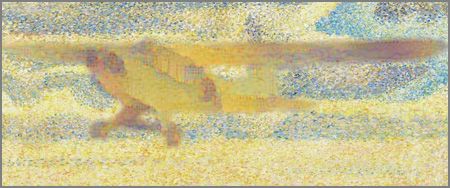}
	\includegraphics[width=.162\textwidth]{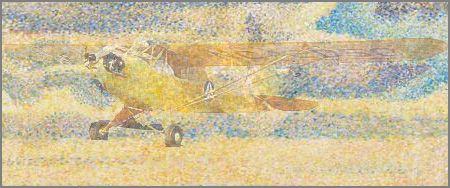}
	\includegraphics[width=.162\textwidth]{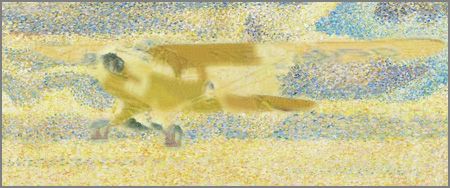}
	\includegraphics[width=.162\textwidth]{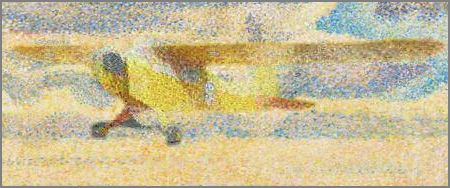}
\\ 
&
	\includegraphics[width=.162\textwidth]{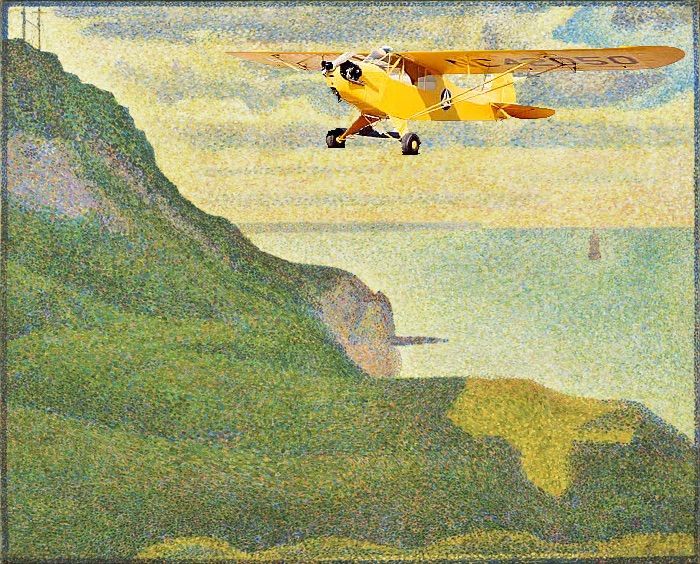}
	\includegraphics[width=.162\textwidth]{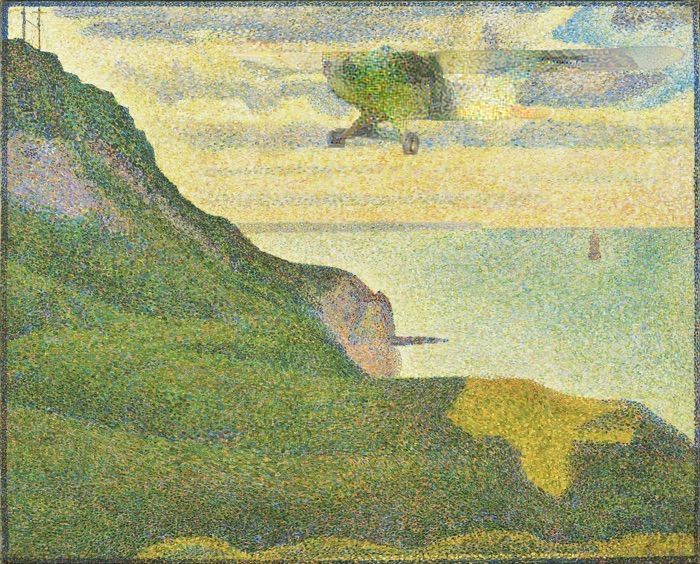}
	\includegraphics[width=.162\textwidth]{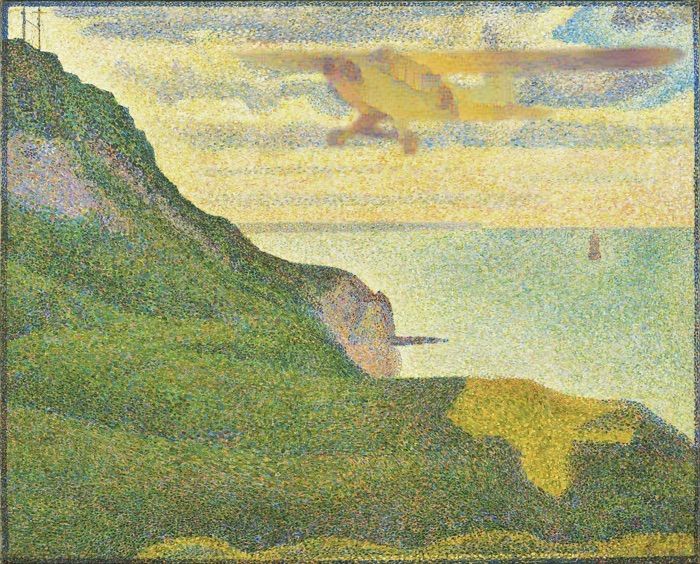}
	\includegraphics[width=.162\textwidth]{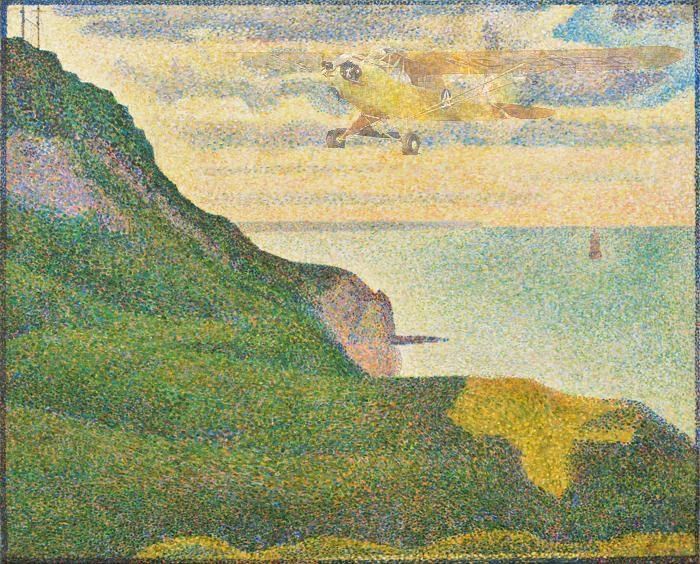}
	\includegraphics[width=.162\textwidth]{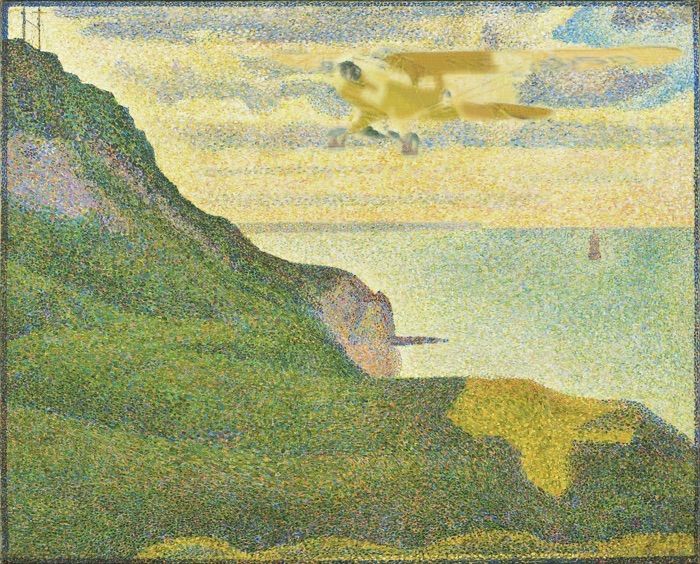}
	\includegraphics[width=.162\textwidth]{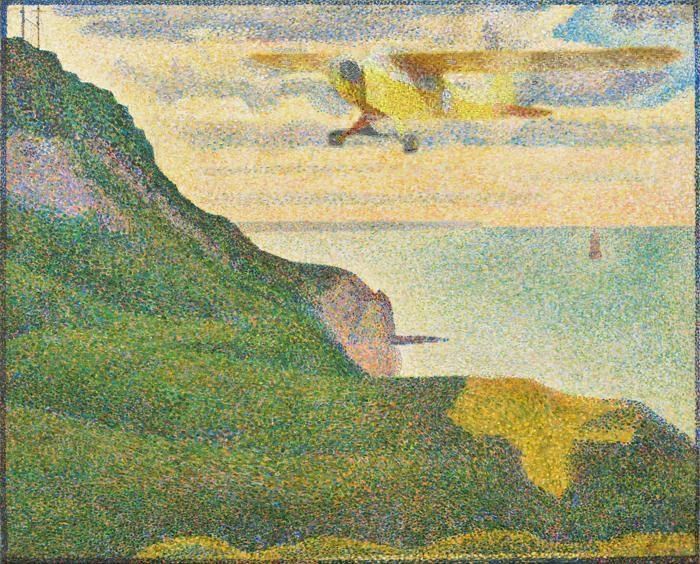}

\\  \hline \\ [-2ex]

\multirow{2}{*}{(vii)}
& 
	\includegraphics[width=.162\textwidth]{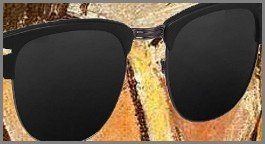}
	\includegraphics[width=.162\textwidth]{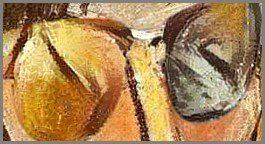}
	\includegraphics[width=.162\textwidth]{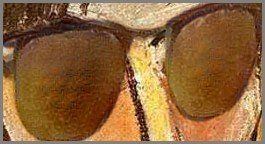}
	\includegraphics[width=.162\textwidth]{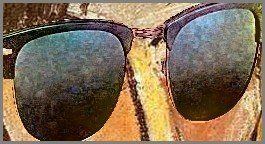}
	\includegraphics[width=.162\textwidth]{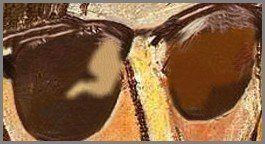}
	\includegraphics[width=.162\textwidth]{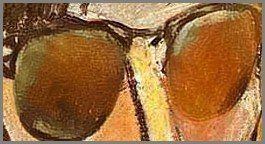}
\\  
&	
	\begin{subfigure}{.162\textwidth}
		\includegraphics[width=1.\linewidth]{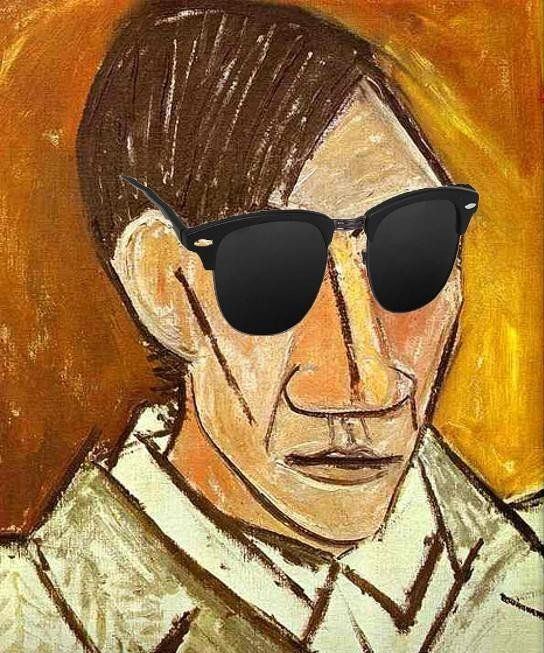}
		\caption{Cut-and-paste}
	\end{subfigure} 
	\begin{subfigure}{.162\textwidth}
		\includegraphics[width=1.\linewidth]{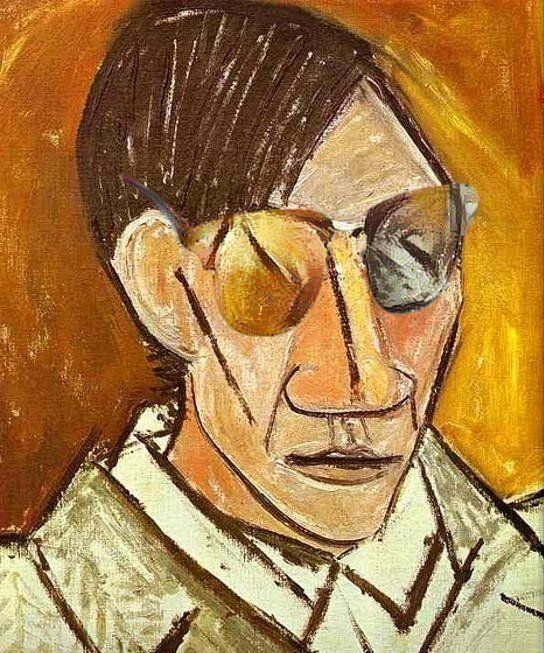}
		\caption{\cite{gatys2015neural}}
	\end{subfigure} 
	\begin{subfigure}{.162\textwidth}
		\includegraphics[width=1.\linewidth]{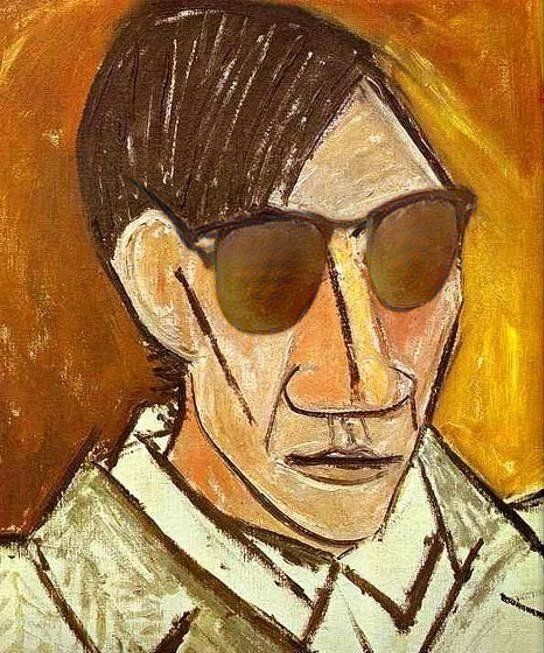}
		\caption{\cite{li2016combining}}
	\end{subfigure}
	\begin{subfigure}{.162\textwidth}
		\includegraphics[width=1.\linewidth]{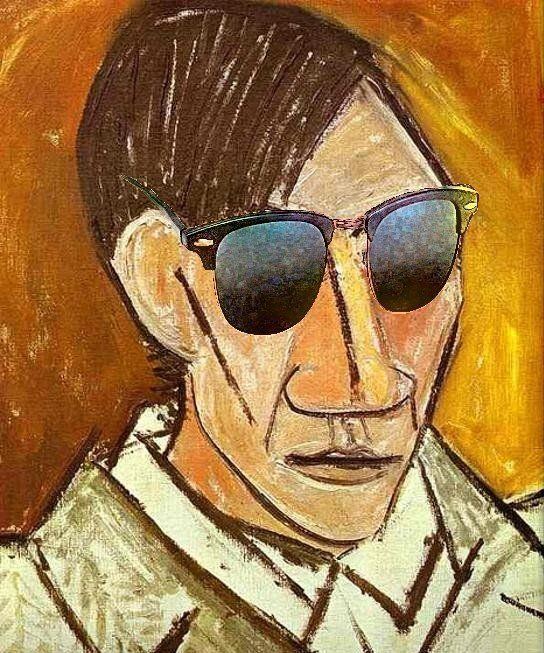}
		\caption{\cite{sunkavalli2010multi}}
	\end{subfigure}
	\begin{subfigure}{.162\textwidth}
		\includegraphics[width=1.\linewidth]{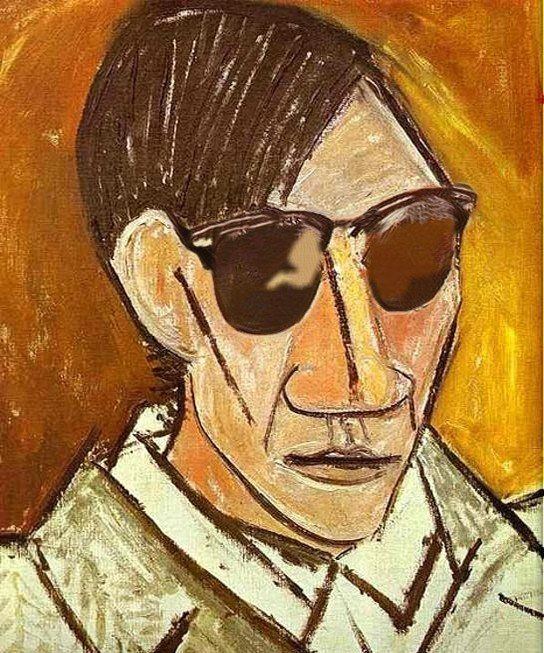}
		\caption{\cite{liao2017visual}}
	\end{subfigure}
	\begin{subfigure}{.162\textwidth}
		\includegraphics[width=1.\linewidth]{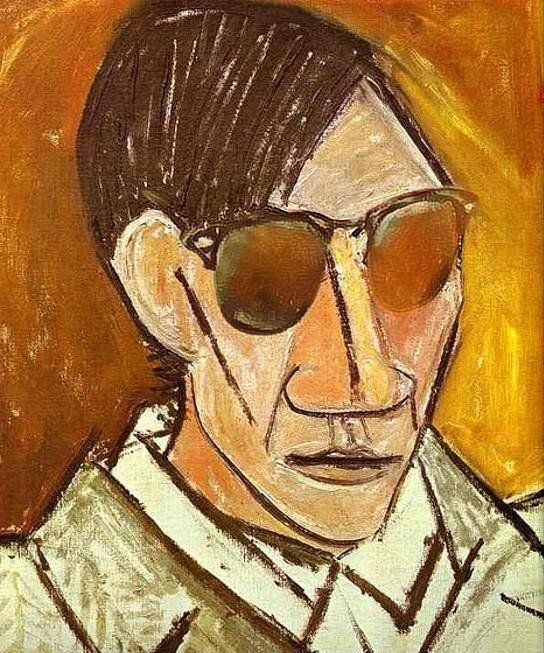}
		\caption{Ours}
	\end{subfigure}

\\   

\end{tabular}

\end{adjustwidth}

\caption{Continued. }

\label{tab:main_res2}

\end{figure*}


\paragraph{Acknowledgements.} This work was supported by a National Science
Foundation grant (IIS-1617861), Adobe, and a Google Faculty Research Award.

\bibliographystyle{eg-alpha-doi}

\bibliography{deeppainting}

\end{document}